\documentclass[apj]{emulateapj}

\def\kmsec{\mbox{km~s$^{\rm -1}$}}
\def\logg{\mbox{log~{\it g}}}
\def\teff{\mbox{$T_{\rm eff}$}}
\def\vt{\mbox{$v_{\rm t}$}}
\def\rpro{\mbox{$r$-process}}
\def\spro{\mbox{$s$-process}}
\def\ncap{\mbox{$n$-capture}}
\def\msun{$M_{\odot}$}
\def\bd{BD$+$17~3248}

\shorttitle{Empirical $s$-process Abundance Distribution}
\shortauthors{Roederer, Marino, \& Sneden}
\submitted{Accepted for publication in the Astrophysical Journal}

\begin{document}

\title{Characterizing the Heavy Elements in Globular Cluster M22
and an Empirical $s$-process Abundance Distribution
Derived from the Two Stellar Groups
\footnotemark[1]}

\footnotetext[1]{
This paper includes data gathered with the 6.5~meter 
Magellan Telescopes located at Las Campanas Observatory, Chile.}

\author{I.~U.\ Roederer,\altaffilmark{2}
A.~F.\ Marino,\altaffilmark{3} and
C.\ Sneden\altaffilmark{4}
}

\altaffiltext{2}{Carnegie Observatories,
813 Santa Barbara Street, Pasadena, CA 91101 USA;
iur@obs.carnegiescience.edu}

\altaffiltext{3}{Max-Planck-Institut f\"{u}r Astrophysik, 
Karl-Schwarzschild-Str.\ 1, 85741 Garching bei M\"{u}nchen, Germany}

\altaffiltext{4}{Department of Astronomy, University of Texas at Austin,
1 University Station, C1400, Austin, TX 78712 USA}

\begin{abstract}

We present an empirical $s$-process abundance distribution
derived with explicit knowledge of the $r$-process component
in the low-metallicity globular cluster M22.
We have obtained high-resolution,
high signal-to-noise spectra for 6 red giants in 
M22 using the MIKE spectrograph on the Magellan-Clay Telescope at
Las Campanas Observatory.
In each star we derive abundances for 44~species of 40~elements, including
24~elements heavier than zinc ($Z =$~30) 
produced by neutron-capture reactions.
Previous studies determined that 3 of these stars 
(the ``$r+s$ group'') have an enhancement of
$s$-process material relative to the other 3~stars (the ``$r$-only group'').
We confirm that the $r+s$ group
is moderately enriched in Pb relative to the $r$-only group.
Both groups of stars were born with the same
amount of $r$-process material, but $s$-process material
was also present in the gas from which the $r+s$ group formed.
The $s$-process abundances are inconsistent with 
predictions for AGB stars with $M \leq$~3~$M_{\odot}$
and suggest an origin in more massive AGB stars capable of
activating the $^{22}$Ne($\alpha$,$n$)$^{25}$Mg reaction.
We calculate the $s$-process ``residual'' by subtracting
the $r$-process pattern in the $r$-only group from the
abundances in the $r+s$ group.
In contrast to previous $r$- and $s$-process decompositions, this
approach makes no assumptions about the $r$- and $s$-process
distributions in the solar system and provides a unique opportunity
to explore $s$-process yields in a metal-poor environment.

\end{abstract}


\section{Introduction}
\label{intro}

By the time the average global star formation rate peaked
in galaxies destined to grow to the size of the
Milky Way 1--3~Gyr after the Big Bang,
vigorous heavy metal enrichment had already begun.  
Elements heavier than the Fe-group
are traditionally understood to be produced
mainly by two processes, the rapid and slow neutron-capture processes.
The \rpro\ (``$r$'' for \textit{rapid}) manufactures heavy nuclei
by overwhelming existing nuclei with a rapid neutron burst
on timescales $\sim$~1~s, far shorter than the average 
$\beta$-decay timescales that could return unstable nuclei
to stable ones.
The \spro\ (``$s$'' for \textit{slow}) manufactures heavy nuclei
by adding neutrons to existing nuclei 
on timescales slow relative to the average $\beta$-decay rates.
Each of these two neutron ($n$) capture processes contributes
about half of the heavy elements in the solar system (S.S.),
which samples the chemistry of the interstellar medium (ISM) 
at one point in the Milky Way disc more than 9~Gyr after the big bang.
The \rpro\ requires an explosive, neutron-rich environment, suggesting an
association with the core collapse supernovae (SNe) that claim the lives of
massive stars ($M \gtrsim$~8~\msun), while the \spro\ 
may be activated in less massive stars (1~$\lesssim M \lesssim$~8~\msun)
during their asymptotic giant branch (AGB) phase of evolution.
Enrichment of the ISM by \rpro\ material may begin a few tens of Myr after 
star formation commences, while \spro\ enrichment requires at least
50~Myr to several Gyr depending on the AGB mass ranges involved.

After an early description of the \spro\ by \citet{burbidge57},
\citet{clayton61} and \citet{seeger65} 
developed the phenomenological (also known as the ``classical'')
approach that dominated
\spro\ modeling for decades to follow.
This method takes advantage of the fact that the product of
the \ncap\ cross section and the \spro\ abundance of each isotope
is slowly variable and can be approximated locally as a constant.
These authors also recognized that a single neutron flux is insufficient
to reproduce the $s$-only isotopes in the S.S.
(See also \citealt{clayton67}.)
In order to explain the \spro\ distribution 
in the S.S., at least 3 components are required, known
today as the ``main,'' ``weak,'' and ``strong'' components.
The main component accounts for 
isotopes from 90~$\lesssim A \leq$~207,
the weak component accounts for the bulk
of the production of isotopes with $A \lesssim$~90,
and the strong component accounts for more than half of $^{208}$Pb.

More and improved experimental data collected over subsequent decades
revealed the shortcomings
of this phenomenological approach (e.g., \citealt*{kappeler89}).
Predictions of \spro\ yields made by post-processing stellar evolution
models with reaction networks 
(e.g., \citealt{arlandini99}) improved the fit, particularly near the
closed neutron shells at $N =$~50, 82, and 126.
Eventually the full reaction networks 
were integrated into the stellar evolution codes
(e.g., \citealt*{straniero06}; \citealt{cristallo09}).
Nucleosynthesis via the \spro\ depends on 
a number of variables including mass, metallicity, 
and \spro\ efficiency.
Uncertainties in the mass dredged up after each thermal
instability (which brings \spro\ material to the surface)
and the mass-loss rate further complicate predictions.
Detailed models are constrained by
spectroscopic observations of \spro\ material in 
intrinsic (i.e., self-enriched) stars or extrinsic (i.e., 
enriched by a binary companion or born with the \spro\ material) ones.
These models have mainly focused on low- and intermediate-mass
AGB stars (i.e., $\leq$~3~\msun) 
and are quite
successful at reproducing both the S.S.\ \spro\ isotopic distribution
and the elemental distributions observed in a variety of stars
(e.g., \citealt{smith86,lambert95,busso99,busso01,bisterzo09,bisterzo11}).

The \spro\ efficiency is largely governed by the
conditions that activate reactions to liberate neutrons,
and two sources of neutrons have been identified in AGB stars.
The first, the $^{13}$C($\alpha$,$n$)$^{16}$O reaction,
is activated at temperatures around 1$\times$10$^{8}$~K.
$^{13}$C is of primary origin, synthesized
from proton captures on freshly produced $^{12}$C.
The $^{13}$C pocket is thought to form in the top layers of
the region between the H and He shell-burning regions
when protons from the H envelope are mixed into this region
during the third dredge up.
The amount of $^{13}$C in the pocket can be thought of as one measure of the
\spro\ efficiency.
The other source, the $^{22}$Ne($\alpha$,$n$)$^{25}$Mg reaction,
is activated at somewhat higher temperatures near 3.5$\times$10$^{8}$~K.
$^{22}$Ne is also primary.
It is produced by the reaction sequence
$^{14}$N($\alpha$,$\gamma$)$^{18}$F($\beta^{+}\nu$)$^{18}$O($\alpha$,$\gamma$)$^{22}$Ne, 
where $^{14}$N is also primary as the most abundant product of CNO burning.
Neutron densities from the $^{13}$C and $^{22}$Ne reactions may reach 
$\sim$~10$^{7}$~cm$^{-3}$ and $\gtrsim$~10$^{11}$~cm$^{-3}$, respectively.
See, e.g., reviews by \citet*{busso99} and \citet{straniero06} 
for further details.

The heavy elements in the S.S.\ are the products
of many and various stars, and
the stellar evolution parameter space necessary to fully reproduce 
the S.S.\ \spro\ pattern is vast and gradually being explored.
Only in the S.S.\ is the complete heavy element inventory 
known with great precision at the isotopic level 
(e.g., \citealt{lodders03}).
Isotopes that can only be formed by the $r$- or \spro\ are 
readily identified, but 
no element in the S.S.\ with 30~$< Z \leq$~83 owes its presence
entirely to the $r$- or \spro.
Limited by the assumption that only two processes contribute,
the $r$- and \spro\ content 
in S.S.\ material can be estimated by the formula
$N_{\odot,r} = N_{\odot,{\rm tot}} - N_{\odot,s}$.
That is, the \rpro\ ``residual'' equals 
the total S.S.\ abundance minus the \spro\ contribution, which is
obtained by either phenomenological or stellar models
(e.g., \citealt{seeger65}, \citealt{cameron73}, \citealt{kappeler89}).
Nearly all abundance information in other stars is in the form of elemental
abundances.
In certain astrophysical environments
only one process or the other contributes,
enabling direct comparison with model predictions.
The difficulty lies in identifying suitable stars 
whose heavy elements may be reliably interpreted as having
originated in only one process or the other.

One such star, \mbox{CS~22892--052}, with a metallicity
less than 1/1000 solar ([Fe/H]~$= -3.1$),\footnote{
We adopt standard definitions of elemental abundances and ratios.
For element X, 
$\log\epsilon$(X)~$\equiv \log_{10}(N_{\rm X}/N_{\rm H}) +$~12.0.
For elements X and Y, 
[X/Y]~$\equiv \log_{10} (N_{\rm X}/N_{\rm Y})_{\star} -
\log_{10} (N_{\rm X}/N_{\rm Y})_{\odot}$.
}
was discovered in the survey of \citet*{beers92}.
\mbox{CS~22892--052}
has a heavy element abundance pattern that very nearly matches
the scaled \rpro\ residuals in the S.S.\ 
(e.g., \citealt{sneden94}, \citealt{cowan95}).
Several other metal-poor stars with this pattern have been found,
and nearly all stars analyzed to
date contain detectable quantities of elements heavier than the Fe-group.
These elements are frequently attributed to \rpro\ nucleosynthesis
(e.g., \citealt{truran81}, \citealt{mcwilliam98}, 
\citealt*{sneden08}, \citealt{roederer10b}).
The consistent \rpro\ abundance pattern observed in several
stars heavily enriched by \rpro\ material inspired the idea 
that \rpro\ abundances everywhere 
(at least for $Z \geq$~56) 
may be scaled versions of the same pattern;
however, stars with less extreme levels of \rpro\ enrichment
clearly deviate from this pattern
(e.g., \citealt{honda07}, \citealt{roederer10b}).

Some metal-poor stars contain \spro\ material mixed
with the \rpro\ contribution. 
Obtaining an empirical measure of the \spro\ content in stars other than the
sun is difficult because a level of \rpro\ enrichment must be assumed.
The metal-poor globular cluster (GC) M22 provides an opportunity to probe
\spro\ enrichment in a low-metallicity environment where the
\rpro\ content is explicitly known.
Recent spectroscopic studies have demonstrated
that star-to-star variations
in heavy elements exist in M22 \citep{marino09,marino11}.
This metal-poor ([Fe/H]~$= -$1.76~$\pm$~0.10)
GC hosts two groups of stars, each with different amounts of
heavy elements (Y, Zr, Ba, La, Nd) that in the S.S.\
are overwhelmingly due to the \spro\ (e.g., \citealt{simmerer04}).
\citeauthor{marino11}\ showed that
the abundances of these elements, together with the total 
C$+$N$+$O and overall Fe-group abundances, increase as a function of
metallicity. 
In contrast, [Eu/Fe] 
has no metallicity dependence
(only 3\% of S.S.\ Eu was produced by the \spro),
demonstrating that the heavy element variations
are due to different amounts of \spro\ material. 
Thus, the chemistry of M22 suggests 
one stellar group formed from gas enriched by \rpro\ 
nucleosynthesis and a second group formed from gas also enriched in
\spro\ material.
Multiple stellar groups in M22 are also revealed
in a split in the sub-giant branch (SGB) revealed by
Hubble Space Telescope photometry \citep{piotto09,marino09}.

\begin{deluxetable*}{ccccccccccccc}
\tablecaption{Photometry, Atmospheric Parameters, Radial Velocities, Exposure Times, and S/N Estimates
\label{phottab}
}
\tablewidth{0pt}
\tablehead{
\colhead{Star} &
\colhead{$V$} &
\colhead{$(B-V)_{0}$} &
\colhead{\teff} &
\colhead{\logg} &
\colhead{\vt} &
\colhead{[Fe/H]} &
\colhead{RV} &
\colhead{$t_{\rm exp}$} &
\colhead{S/N} &
\colhead{S/N} &
\colhead{S/N} &
\colhead{S/N} \\
\colhead{} &
\colhead{} &
\colhead{} &
\colhead{(K)} &
\colhead{} &
\colhead{(\kmsec)} &
\colhead{} &
\colhead{(\kmsec)} &
\colhead{(s)} &
\colhead{(3950\AA)} &
\colhead{(4550\AA)} &
\colhead{(5200\AA)} &
\colhead{(6750\AA)} }
\startdata
I-27   & 12.39 & 1.28 & 4455 & 1.45 & 1.60 & $-$1.73 & $-$127.8 & 2600 & 40/1 &  95/1 &  95/1 & 210/1 \\
I-37   & 12.01 & 1.45 & 4370 & 1.05 & 1.50 & $-$1.73 & $-$157.7 & 3800 & 50/1 & 125/1 & 140/1 & 340/1 \\
I-53   & 12.69 & 1.36 & 4500 & 1.35 & 1.55 & $-$1.74 & $-$145.4 & 3000 & 45/1 & 105/1 & 120/1 & 270/1 \\
I-80   & 12.53 & 1.38 & 4460 & 1.15 & 1.55 & $-$1.70 & $-$149.8 & 3100 & 40/1 &  90/1 & 100/1 & 230/1 \\
III-33 & 12.25 & 1.40 & 4430 & 1.05 & 1.70 & $-$1.78 & $-$145.8 & 1600 & 45/1 & 105/1 & 120/1 & 275/1 \\
IV-59  & 11.93 & 1.45 & 4400 & 1.00 & 1.70 & $-$1.77 & $-$152.8 & 1600 & 45/1 & 110/1 & 125/1 & 300/1 \\
\enddata
\end{deluxetable*}

The chemical pattern revealed in M22 makes this GC a suitable target to
investigate \spro\ abundance distributions.
Observations indicate that
the \rpro\ content of both stellar groups in M22 is the same.
Observations also indicate that
the more metal-rich group 
(hereafter referred to as the ``$r+s$ group'')
was formed from gas also enriched by \spro\ material.
We can subtract the \rpro\ abundance pattern
(established empirically in the metal-poor group,
hereafter referred to as the ``$r$-only group'') from the
abundance pattern in the $r+s$ group 
to derive an empirical \spro\ abundance distribution.
One favorable aspect of this approach is that it does not rely on the
decomposition of S.S.\ material into $r$- or \spro\ fractions
to interpret abundances elsewhere in the Galaxy.

\section{Observations}
\label{observations}

Six probable members of M22 were observed with the 
Magellan Inamori Kyocera Echelle (MIKE)
spectrograph \citep{bernstein03} on the
6.5~m Magellan-Clay Telescope at Las Campanas Observatory
on 2011 March 17--18.
These spectra were taken with the 0.7''$\times$5.0'' slit yielding
a resolving power of $R \sim$~41,000 in the blue 
and $R \sim$~35,000 in the red, split by a dichroic
around 4950\AA.
This setup provides complete wavelength coverage from 
3350--9150\AA, though in practice we only make use of the
region from 3690 to 7800\AA\ where the lines of interest are located.
Data reduction, extraction, and wavelength calibration were performed using 
the MIKE data reduction pipeline
written by D.\ Kelson.
(See also \citealt{kelson03}).
Continuum normalization and order stitching were performed within the 
IRAF environment.\footnote{
IRAF is distributed by the National Optical Astronomy Observatories,
which are operated by the Association of Universities for Research
in Astronomy, Inc., under cooperative agreement with the National
Science Foundation.}

The six observed stars are all cool giants on the M22 red giant branch (RGB).
Table~\ref{phottab} lists the photometry from the 
Stetson database (priv.\ communication,
corrected for differential reddening as in \citealt{marino11})
atmospheric parameters (adopted from \citeauthor{marino11};
see Section~\ref{analysis}), 
Heliocentric radial velocities (RV), exposure times, and
signal-to-noise (S/N) estimates for our targets.
We estimate the S/N based on
Poisson statistics for the number of photons collected 
in the continuum.
We measure the RV with respect to the ThAr lamp 
by cross correlating the echelle order 
containing the Mg~\textsc{i} \textit{b} lines in each spectrum
against a template
using the \textit{fxcor} task in IRAF.
We create the template by measuring the wavelengths of
unblended Fe~\textsc{i} lines in this order in star
\mbox{IV-59}, which has the highest S/N in a single exposure.
We compute velocity corrections to the Heliocentric rest frame
using the IRAF \textit{rvcorrect} task.
This method yields a total uncertainty of 0.8~\kmsec\ 
per observation (see \citealt{roederer10a}).
Our RVs are in good agreement ($\Delta =$~0.7~$\pm$~0.7~\kmsec)
with those derived by \citet{marino09} for 4 stars in common.

\section{Abundance Analysis}
\label{analysis}

We perform a standard abundance analysis on the 6~stars observed
with MIKE.
We adopt the atmospheric parameters derived by \citet{marino11} and use
$\alpha$-enhanced
ATLAS9 model atmospheres from \citet{castelli04}.
We perform the analysis using the latest version of the
spectral analysis code MOOG \citep{sneden73},
with updates to the calculation of the
Rayleigh scattering contribution to the
continuous opacity described in \citet{sobeck11}.
We measure equivalent widths (EWs) by fitting Voigt absorption
line profiles to the continuum-normalized spectra, and we
derive abundances of 
Na~\textsc{i}, 
Mg~\textsc{i}, 
Al~\textsc{i}, 
Si~\textsc{i},
K~\textsc{i},
Ca~\textsc{i},
Ti~\textsc{i} and \textsc{ii},
Cr~\textsc{i} and \textsc{ii},
Fe~\textsc{i} and \textsc{ii},
Ni~\textsc{i}, and
Zn~\textsc{i} from a standard EW analysis.
Abundances of all other elements are derived by spectral synthesis,
comparing synthetic spectra to the observations.
This is necessary for species whose lines 
may be blended, have broad hyperfine structure (HFS),
or have multiple isotopes whose electronic levels are shifted slightly.
The linelist, atomic data and references, 
and derived abundances for each line
are presented in Table~\ref{lineabundtab}, which is available 
in the online-only edition.

\begin{figure*}
\begin{center}
\includegraphics[angle=0,width=2.8in]{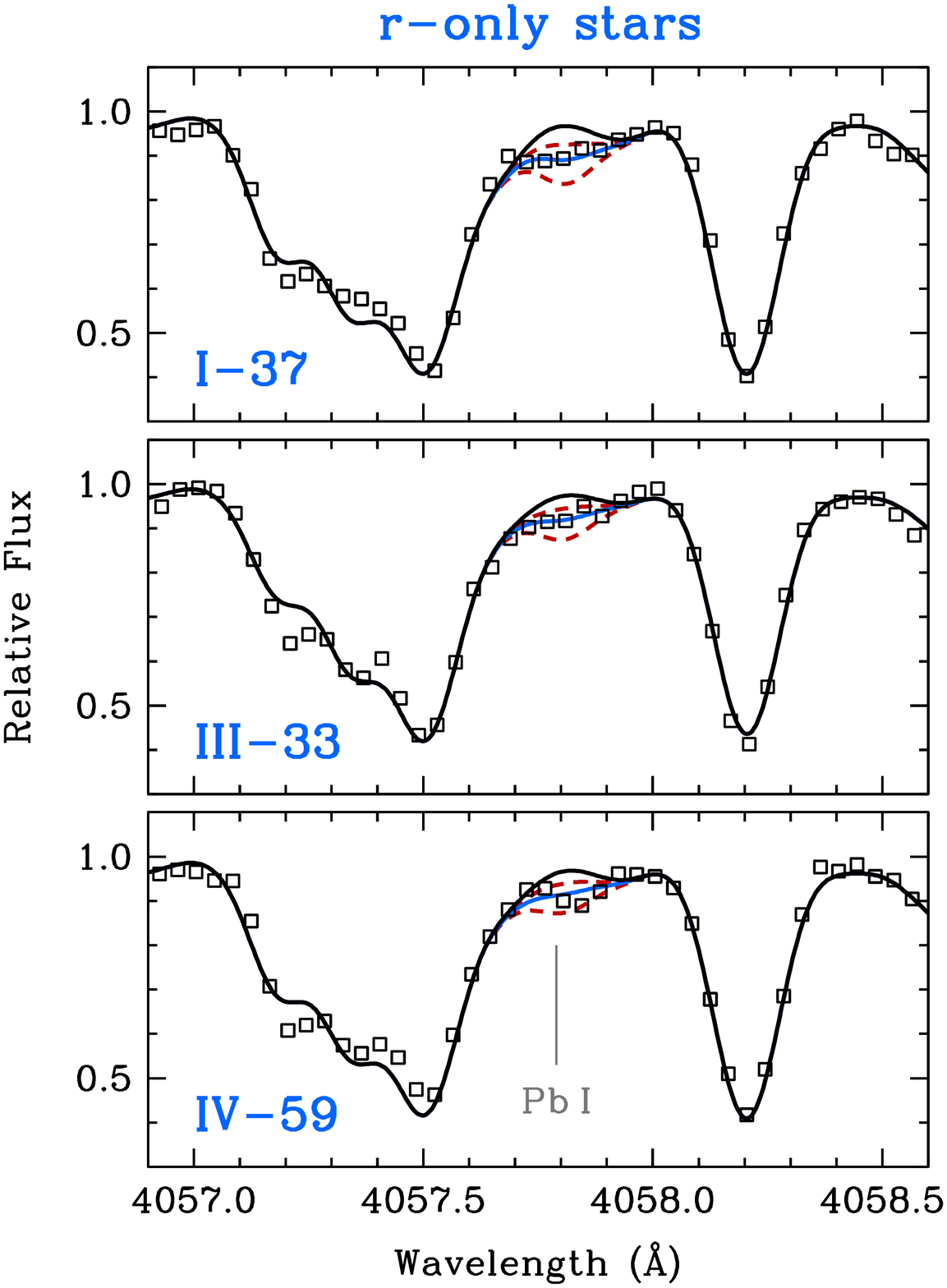} 
\hspace*{0.1in}
\includegraphics[angle=0,width=2.8in]{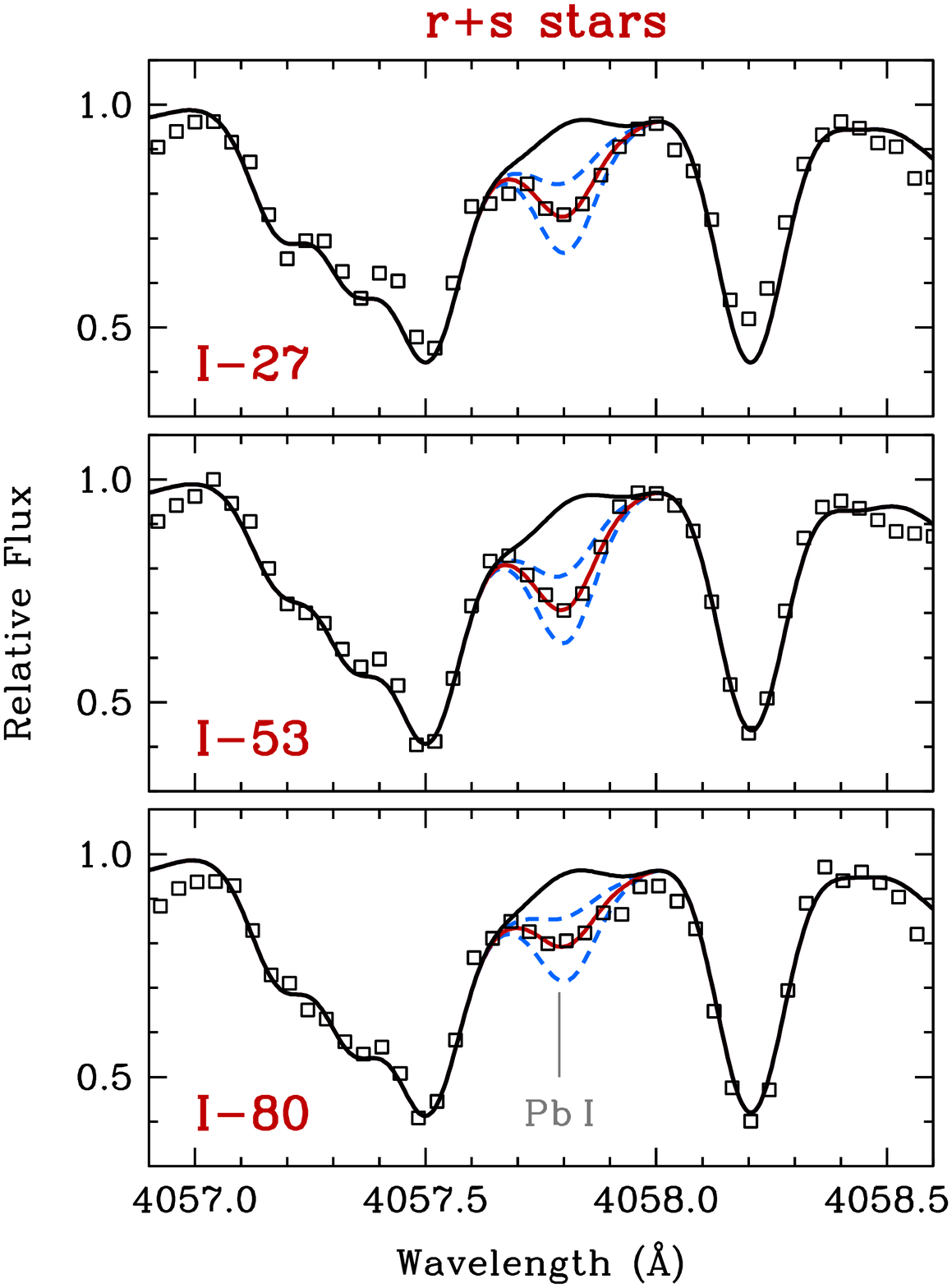}
\end{center}
\caption{
\label{pbplot}
Comparison of observed (open squares) and synthetic (lines) spectra
around the Pb~\textsc{i} 4057.8\AA\ line.
The left panels show the 3 $r$-only stars, and the
right panels show the 3 $r+s$ stars.
The solid colored line (left panels, blue; right panels, red)
indicates the best-fit abundance, 
the dashed lines indicate variations in the best-fit abundance
by 0.3~dex, and the
solid black line indicates a synthesis with no Pb~\textsc{i} present.
}
\end{figure*}

\begin{deluxetable*}{ccccccccccc}
\tablecaption{Line-by-line Abundances
\label{lineabundtab}
}
\tablewidth{0pt}
\tablehead{
\colhead{Species} &
\colhead{$\lambda$} &
\colhead{E.P.} &
\colhead{$\log(gf)$} &
\colhead{Ref.} &
\colhead{I-37} &
\colhead{III-33} &
\colhead{IV-59} &
\colhead{I-27} &
\colhead{I-53} &
\colhead{I-80} \\
\colhead{} &
\colhead{(\AA)} &
\colhead{(eV)} &
\colhead{} &
\colhead{} &
\colhead{} &
\colhead{} &
\colhead{} &
\colhead{} &
\colhead{} &
\colhead{} }
\startdata
 Na~\textsc{i}  & 5682.63 & 2.10 & $-$0.71 &  1  &  4.68        & 4.09    & 4.95    & 4.69    & 4.70    & 5.10    \\
 Na~\textsc{i}  & 5688.20 & 2.10 & $-$0.45 &  1  &  4.81        & 4.19    & 5.05    & 4.88    & 4.81    & 5.18    \\
 Na~\textsc{i}  & 6154.23 & 2.10 & $-$1.55 &  1  &  4.63        & 4.07    & 4.98    & \nodata & 4.63    & 5.04    \\
 Na~\textsc{i}  & 6160.75 & 2.10 & $-$1.25 &  1  &  4.77        & 4.13    & 4.89    & 4.68    & 4.67    & 5.13    \\
 \enddata
\tablerefs{
(1) \citet{fuhr09};
(2) \citet{chang90};
(3) \citet{lawler89}, using HFS from \citet{kurucz95};
(4) \citet{blackwell82a,blackwell82b}, increased by 0.056~dex according to \citet{grevesse89};
(5) \citet{pickering01}, with corrections given in \citet{pickering02};
(6) \citet{whaling85}, using HFS from \citet{kurucz95};
(7) \citet{sobeck07};
(8) \citet{nilsson06};
(9) \citet{blackwellwhitehead07}, using HFS from \citet{kurucz95};
(10) \citet{booth84}, using HFS from \citet{kurucz95}
(11) \citet{obrian91};
(12) \citet{melendez09};
(13) \citet{nitz99}, using HFS from \citet{kurucz95};
(14) \citet{cardon82}, using HFS from \citet{kurucz95};
(15) \citet{wickliffe97a};
(16) \citet{bielski75}, using HFS from \citet{kurucz95};
(17) \citet{biemont80};
(18) \citet{migdalek87};
(19) \citet{hannaford82};
(20) \citet{biemont81};
(21) \citet{ljung06};
(22) \citet{whaling88}; 
(23) \citet{wickliffe94};
(24) \citet{duquette85};
(25) \citet{fuhr09}, using HFS from \citet{mcwilliam98};
(26) \citet{lawler01a}, using HFS from \citet{ivans06};
(27) \citet{lawler09}; 
(28) \citet{li07}, using HFS from \citet{sneden09};
(29) \citet{ivarsson01}, using HFS from \citet{sneden09};
(30) \citet{denhartog03};
(31) \citet{lawler06};
(32) \citet{lawler01b}, using HFS from \citet{ivans06};
(33) \citet{denhartog06};
(34) \citet{lawler01c}, using HFS from \citet{lawler01d};
(35) \citet{wickliffe00}; 
(36) \citet{lawler04} for both $\log(gf)$ value and HFS;
(37) \citet{lawler08};
(38) \citet{wickliffe97b};
(39) \citet{sneden09} for both $\log(gf)$ value and HFS;
(40) \citet{lawler07};
(41) \citet{ivarsson03}; 
(42) \citet{biemont00}; 
(43) \citet{nilsson02}. 
}
\tablecomments{
The full version of this table is available in the online edition of the journal.
A limited version is given here to indicate the form and content of the data.
Abundances are given as $\log\epsilon$ notation.  
A ``:'' indicates the derived abundance is less secure,
and we estimate an uncertainty of 0.2~dex.
See text for details.
}
\end{deluxetable*}

In Figure~\ref{pbplot} we show synthetic spectra fits to the
region around the Pb~\textsc{i} line at 4057\AA.
The solid lines represent the best-fit abundance, and the dashed
lines represent variations in this fit by 0.3~dex.
The 6~stars in our sample have very similar atmospheric parameters.
In our syntheses we adjust the linelist to fit blending features in
one star and leave these adjustments unchanged in the analysis of
other stars.
This preserves a differential quality in the abundance analysis,
which is important in the case of
abundances derived from very few or heavily blended lines. 

\begin{deluxetable}{ccccc}
\tablecaption{Effect of Different Analysis Tools on Derived Abundances in \mbox{I-37}
\label{lacetab}
}
\tablewidth{0pt}
\tablehead{
\colhead{$\lambda$ (\AA)} &
\colhead{$\log\epsilon$\tablenotemark{A}} &
\colhead{$\log\epsilon$\tablenotemark{B}} &
\colhead{$\log\epsilon$\tablenotemark{C}} &
\colhead{$\log\epsilon$\tablenotemark{D}} }
\startdata
\multicolumn{5}{c}{La~\textsc{ii} Lines} \\
\hline
3988.51  & $-$0.74  & $-$0.66  & $-$0.71  & $-$0.61 \\
3995.74  & $-$0.70  & $-$0.56  & $-$0.69  & $-$0.51 \\
4086.71  & $-$0.78  & $-$0.70  & $-$0.70  & $-$0.56 \\
4322.50  & $-$0.70  & $-$0.67  & $-$0.68  & $-$0.66 \\
4662.50  & $-$0.54  & $-$0.50  & $-$0.55  & $-$0.46 \\
4748.73  & $-$0.65  & $-$0.65  & $-$0.62  & $-$0.70 \\
4804.04  & $-$0.44  & $-$0.43  & $-$0.43  & $-$0.53 \\
4920.98  & $-$0.34  & $-$0.31  & $-$0.28  & $-$0.25 \\
4986.82  & $-$0.49  & $-$0.48  & $-$0.47  & $-$0.41 \\
5114.56  & $-$0.48  & $-$0.45  & $-$0.43  & $-$0.42 \\
5290.84  & $-$0.73  & $-$0.69  & $-$0.71  & $-$0.64 \\
5303.53  & $-$0.45  & $-$0.46  & $-$0.44  & $-$0.45 \\
6262.29  & $-$0.44  & $-$0.45  & $-$0.41  & $-$0.41 \\
6390.48  & $-$0.39  & $-$0.42  & $-$0.38  & $-$0.40 \\
\hline
Mean:              & $-$0.56 & $-$0.53 & $-$0.54 & $-$0.50 \\
$\sigma$:          & 0.15    & 0.12    & 0.15    & 0.12    \\   
$\sigma/\sqrt{N}$: & 0.040   & 0.033   & 0.038   & 0.033   \\
\hline   
\hline
\multicolumn{5}{c}{Ce~\textsc{ii} Lines} \\
\hline
4073.47  & $-$0.35  & $-$0.24  & $-$0.32  & $-$0.23 \\
4083.22  & $-$0.25  & $-$0.17  & $-$0.21  & $-$0.14 \\
4120.83  & $-$0.20  & $-$0.17  & $-$0.17  & $-$0.08 \\
4127.36  & $-$0.24  & $-$0.20  & $-$0.24  & $-$0.15 \\
4137.65  & $-$0.26  & $-$0.18  & $-$0.20  & $-$0.13 \\
4222.60  & $-$0.23  & $-$0.19  & $-$0.20  & $-$0.16 \\
4364.65  & $-$0.21  & $-$0.18  & $-$0.19  & $-$0.15 \\
4418.78  & $-$0.23  & $-$0.15  & $-$0.16  & $-$0.08 \\
4486.91  & $-$0.26  & $-$0.24  & $-$0.24  & $-$0.23 \\
4560.96  & $-$0.21  & $-$0.19  & $-$0.18  & $-$0.17 \\
4562.36  & $-$0.17  & $-$0.14  & $-$0.14  & $-$0.08 \\
4572.28  & $-$0.16  & $-$0.11  & $-$0.02  & $+$0.00 \\ 
4582.50  & $-$0.05  & $-$0.02  & $-$0.05  & $+$0.03 \\ 
4628.16  & $-$0.08  & $-$0.07  & $-$0.02  & $+$0.04 \\ 
5274.23  & $-$0.20  & $-$0.21  & $-$0.22  & $-$0.20 \\
5330.56  & $-$0.22  & $-$0.24  & $-$0.19  & $-$0.22 \\
\hline
Mean:              & $-$0.21 & $-$0.17 & $-$0.17 & $-$0.12 \\
$\sigma$:          & 0.07    & 0.06    & 0.08    & 0.09    \\   
$\sigma/\sqrt{N}$: & 0.017   & 0.015   & 0.020   & 0.021   \\
\hline
\hline
$\log\epsilon$(La/Ce):& $-$0.35    & $-$0.36    & $-$0.36    & $-$0.38    \\
                      & $\pm$0.043 & $\pm$0.036 & $\pm$0.044 & $\pm$0.039 \\
\enddata
\tablenotetext{A}{MOOG with scattering, MARCS model}
\tablenotetext{B}{MOOG without scattering, MARCS model}
\tablenotetext{C}{MOOG with scattering, ATLAS9 model}
\tablenotetext{D}{MOOG without scattering, ATLAS9 model}
\end{deluxetable}

The EWs measured by \citet{marino11} from a high S/N VLT/UVES spectrum
of \mbox{I-27} are systematically higher by 4.9~$\pm$~0.6~m\AA\
($\sigma =$~3.2~m\AA).
the only case where the offset is larger than the standard deviation
of the residuals (3.2--3.4~m\AA\ in all stars).
In \mbox{I-27}, this translates to no significant difference
in the derived [X/Fe] ratios (since both abundances are 
similarly affected), $\Delta$[X/Fe]~$=$~0.00~$\pm$~0.01 
($\sigma =$~0.03~dex),
and the change in metallicity is $\Delta$[Fe/H]~$=$~0.06.
EWs measured from 
the lower S/N APO/ARC Echelle spectrum of \mbox{I-27} analyzed by
\citeauthor{marino11}\ are systematically lower by 5.2~$\pm$~2.6~m\AA\
($\sigma =$~11~m\AA).
Differences at this level may be expected when analyzing spectra 
of various resolution and S/N, collected over many nights 
using instruments
on different telescopes in different hemispheres,
so we do not pursue the matter further.
We also compare our derived abundance ratios with those
presented in \citeauthor{marino11} 
After accounting for the different sets of $\log(gf)$ values,
S.S.\ abundances (see \citealt{marino09}), 
and line-by-line mean offsets (see Appendix),
the abundance offsets
for most [Fe/H] and [X/Fe] ratios can be immediately accounted for.
Offsets in other species 
(Ti~\textsc{i}, Cu~\textsc{i}, Zn~\textsc{i}, 
La~\textsc{ii}, and Nd~\textsc{ii})
are unexplained by these factors but probably result from 
the S/N and small numbers of features available in the
spectra of \citeauthor{marino11} 
For the purposes of the present study, we will focus on the
internal abundance differences derived from our MIKE spectra.

Table~\ref{lacetab} shows the internal abundance precision possible
with this method when large numbers ($N >$~10) of lines are available
across the visible spectral range.
For this test we derive the abundances of La~\textsc{ii} and Ce~\textsc{ii}
in star \mbox{I-37} using two grids of model atmospheres
(MARCS, \citealt{gustafsson08}; ATLAS9, \citealt{castelli04})
and different treatments of Rayleigh scattering in MOOG.
The elemental abundances and ratios 
are not dependent on the choice of model atmosphere grids or 
the treatment of Rayleigh scattering.
For example, as shown in Table~\ref{lacetab}, 
in response to the different analysis tools the derived
$\log\epsilon$(La/Ce) ratio changes by no more than 0.03~dex,
which is smaller than the statistical uncertainties
(0.036 to 0.044~dex).
Furthermore, the stars in our study were chosen to have similar 
colors (1.36~$\leq (B-V)_{0} \leq$~1.52), 
metallicities ($-$1.80~$\leq$~[Fe/H]~$\leq -$1.70), 
and atmospheric parameters
(4370~$\leq \teff \leq$~4500~K and 1.00~$\leq \logg \leq$~1.45;
all based on the values presented in \citealt{marino11}).\footnote{
For comparison, precision abundance analyses of nearby
metal-rich dwarfs with stellar parameters similar to the sun often
consider stars with \teff\ within 100~K, \logg\ within 0.1~dex, and
[Fe/H] within 0.1~dex of the solar values to be ``solar twins'' 
(e.g., \citealt{ramirez09}).  }
Thus a relative abundance analysis is appropriate, 
and in all subsequent discussion, tables, and figures
we cite internal (i.e., observational) uncertainties only.

\begin{deluxetable*}{ccccccccccccccccccc}
\tablecaption{Mean Abundances in the Three $r$-only Stars
\label{rabundtab}
}
\tablewidth{0pt}
\tablehead{
\colhead{} &
\colhead{} &
\multicolumn{5}{c}{I-37} &
\colhead{} &
\multicolumn{5}{c}{III-33} &
\colhead{} &
\multicolumn{5}{c}{IV-59} \\
\cline{3-7} \cline{9-13} \cline{15-19}
\colhead{Species} &
\colhead{$Z$} &
\colhead{$\langle\log\epsilon\rangle$} &
\colhead{$\langle$[X/Fe]$\rangle$} &
\colhead{$N$} &
\colhead{$\sigma$} &
\colhead{$\sigma_{\mu}$} &
\colhead{} &
\colhead{$\langle\log\epsilon\rangle$} &
\colhead{$\langle$[X/Fe]$\rangle$} &
\colhead{$N$} &
\colhead{$\sigma$} &
\colhead{$\sigma_{\mu}$} &
\colhead{} &
\colhead{$\langle\log\epsilon\rangle$} &
\colhead{$\langle$[X/Fe]$\rangle$} &
\colhead{$N$} &
\colhead{$\sigma$} &
\colhead{$\sigma_{\mu}$}
}
\startdata
Na~\textsc{i}  & 11 & 4.72  & 0.20  & 4  & 0.08  & 0.041 &  & 4.12  & $-$0.26 & 4  & 0.05  & 0.026 &  & 4.97  & 0.53  & 4  & 0.07  & 0.033 \\
Mg~\textsc{i}  & 12 & 6.25  & 0.37  & 3  & 0.21  & 0.121 &  & 6.17  & 0.43  & 3  & 0.17  & 0.101 &  & 6.16  & 0.36  & 3  & 0.20  & 0.115 \\
Al~\textsc{i}  & 13 & 4.89  & 0.16  & 4  & 0.09  & 0.047 &  & 4.50  & $-$0.09 & 1  & 0.11  & 0.110 &  & 5.33  & 0.68  & 4  & 0.11  & 0.053 \\
Si~\textsc{i}  & 14 & 5.89  & 0.10  & 3  & 0.24  & 0.140 &  & 5.80  & 0.16  & 3  & 0.17  & 0.098 &  & 5.86  & 0.15  & 3  & 0.18  & 0.104 \\
K~\textsc{i}   & 19 & 4.12  & 0.81  & 1  & 0.11  & 0.110 &  & 3.93  & 0.76  & 1  & 0.11  & 0.110 &  & 4.06  & 0.83  & 1  & 0.11  & 0.110 \\
Ca~\textsc{i}  & 20 & 5.04  & 0.41  & 8  & 0.11  & 0.038 &  & 4.84  & 0.36  & 8  & 0.15  & 0.054 &  & 4.93  & 0.40  & 8  & 0.10  & 0.035 \\
Sc~\textsc{ii} & 21 & 1.68  & 0.15  & 5  & 0.27  & 0.120 &  & 1.58  & 0.20  & 5  & 0.30  & 0.134 &  & 1.59  & 0.28  & 5  & 0.29  & 0.128 \\
Ti~\textsc{i}  & 22 & 3.28  & 0.05  & 9  & 0.07  & 0.022 &  & 3.11  & 0.03  & 9  & 0.07  & 0.022 &  & 3.25  & 0.11  & 9  & 0.08  & 0.027 \\
Ti~\textsc{ii} & 22 & 3.79  & 0.47  & 9  & 0.12  & 0.039 &  & 3.58  & 0.40  & 9  & 0.10  & 0.035 &  & 3.64  & 0.53  & 9  & 0.09  & 0.031 \\
V~\textsc{i}   & 23 & 2.00  & $-$0.21 & 5  & 0.10  & 0.045 &  & 1.97  & $-$0.10 & 5  & 0.12  & 0.052 &  & 2.02  & $-$0.11 & 5  & 0.09  & 0.040 \\
Cr~\textsc{i}  & 24 & 3.74  & $-$0.18 & 6  & 0.12  & 0.049 &  & 3.62  & $-$0.15 & 6  & 0.08  & 0.033 &  & 3.72  & $-$0.12 & 6  & 0.07  & 0.029 \\
Cr~\textsc{ii} & 24 & 4.14  & 0.13  & 1  & 0.11  & 0.110 &  & 4.00  & 0.13  & 1  & 0.11  & 0.110 &  & 3.95  & 0.15  & 1  & 0.11  & 0.110 \\
Mn~\textsc{i}  & 25 & 3.13  & $-$0.59 & 4  & 0.12  & 0.061 &  & 3.12  & $-$0.45 & 4  & 0.12  & 0.058 &  & 3.16  & $-$0.47 & 4  & 0.12  & 0.061 \\
Fe~\textsc{i}  & 26 & 5.78  & $-$1.72 & 69 & 0.13  & 0.016 &  & 5.63  & $-$1.87 & 69 & 0.10  & 0.012 &  & 5.70  & $-$1.80 & 69 & 0.12  & 0.015 \\
Fe~\textsc{ii} & 26 & 5.87  & $-$1.63 & 7  & 0.17  & 0.062 &  & 5.73  & $-$1.77 & 7  & 0.08  & 0.031 &  & 5.66  & $-$1.84 & 7  & 0.09  & 0.034 \\
Co~\textsc{i}  & 27 & 3.11  & $-$0.16 & 4  & 0.09  & 0.043 &  & 3.01  & $-$0.12 & 4  & 0.13  & 0.065 &  & 3.06  & $-$0.13 & 4  & 0.18  & 0.089 \\
Ni~\textsc{i}  & 28 & 4.28  & $-$0.23 & 10 & 0.10  & 0.033 &  & 4.20  & $-$0.16 & 10 & 0.12  & 0.039 &  & 4.22  & $-$0.19 & 10 & 0.11  & 0.034 \\
Cu~\textsc{i}  & 29 & 1.83  & $-$0.65 & 2  & 0.08  & 0.057 &  & 1.64  & $-$0.69 & 2  & 0.08  & 0.057 &  & 1.74  & $-$0.65 & 2  & 0.08  & 0.057 \\
Zn~\textsc{i}  & 30 & 2.89  & 0.04  & 2  & 0.08  & 0.055 &  & 2.75  & 0.05  & 2  & 0.08  & 0.057 &  & 2.80  & 0.04  & 2  & 0.08  & 0.057 \\
Rb~\textsc{i}  & 37 & $<$1.00 & $<$0.20 & 1  &\nodata&\nodata&  & $<$1.10 & $<$0.44 & 1  &\nodata&\nodata&  & $<$1.10 & $<$0.38 & 1  &\nodata&\nodata\\
Sr~\textsc{i}  & 38 & 0.60  & $-$0.55 & 1  & 0.11  & 0.110 &  & 0.51  & $-$0.50 & 1  & 0.11  & 0.110 &  & 0.54  & $-$0.53 & 1  & 0.11  & 0.110 \\
Y~\textsc{ii}  & 39 & 0.34  & $-$0.25 & 7  & 0.04  & 0.016 &  & 0.18  & $-$0.26 & 7  & 0.07  & 0.026 &  & 0.18  & $-$0.19 & 7  & 0.08  & 0.030 \\
Zr~\textsc{i}  & 40 & 0.69  & $-$0.18 & 2  & 0.08  & 0.057 &  & 0.78  & 0.06  & 2  & 0.08  & 0.057 &  & 0.63  & $-$0.15 & 2  & 0.12  & 0.085 \\
Zr~\textsc{ii} & 40 & 1.06  & 0.11  & 3  & 0.09  & 0.054 &  & 1.04  & 0.23  & 3  & 0.09  & 0.050 &  & 1.05  & 0.31  & 3  & 0.09  & 0.050 \\
Mo~\textsc{i}  & 42 & 0.33  & 0.16  & 2  & 0.14  & 0.099 &  & 0.00  & $-$0.02 & 1  & 0.20  & 0.200 &  & 0.12  & 0.04  & 3  & 0.28  & 0.164 \\
Ru~\textsc{i}  & 44 & 0.09  & 0.06  & 3  & 0.14  & 0.082 &  & 0.07  & 0.18  & 3  & 0.13  & 0.075 &  & 0.18  & 0.23  & 1  & 0.20  & 0.200 \\
Rh~\textsc{i}  & 45 & $-$0.84 & $-$0.18 & 1  & 0.11  & 0.110 &  &\nodata&\nodata& 0  &\nodata&\nodata&  & $-$1.16 & $-$0.42 & 1  & 0.11  & 0.110 \\
Ba~\textsc{ii} & 56 & 0.76  & 0.20  & 2  & 0.08  & 0.057 &  & 0.53  & 0.12  & 2  & 0.15  & 0.105 &  & 0.53  & 0.19  & 2  & 0.18  & 0.130 \\
La~\textsc{ii} & 57 & $-$0.54 & $-$0.01 & 14 & 0.14  & 0.038 &  & $-$0.59 & 0.09  & 14 & 0.08  & 0.022 &  & $-$0.61 & 0.13  & 14 & 0.11  & 0.029 \\
Ce~\textsc{ii} & 58 & $-$0.17 & $-$0.13 & 16 & 0.08  & 0.020 &  & $-$0.26 & $-$0.06 & 16 & 0.05  & 0.013 &  & $-$0.29 & $-$0.03 & 16 & 0.07  & 0.018 \\
Pr~\textsc{ii} & 59 & $-$0.87 & 0.04  & 4  & 0.05  & 0.025 &  & $-$0.87 & 0.18  & 4  & 0.08  & 0.040 &  & $-$0.93 & 0.19  & 4  & 0.05  & 0.025 \\
Nd~\textsc{ii} & 60 & $-$0.13 & 0.07  & 24 & 0.08  & 0.017 &  & $-$0.18 & 0.18  & 24 & 0.07  & 0.014 &  & $-$0.22 & 0.20  & 24 & 0.08  & 0.016 \\
Sm~\textsc{ii} & 62 & $-$0.48 & 0.19  & 9  & 0.07  & 0.022 &  & $-$0.51 & 0.30  & 9  & 0.07  & 0.022 &  & $-$0.55 & 0.33  & 9  & 0.07  & 0.023 \\
Eu~\textsc{ii} & 63 & $-$0.85 & 0.26  & 3  & 0.20  & 0.113 &  & $-$0.88 & 0.37  & 3  & 0.15  & 0.085 &  & $-$0.92 & 0.40  & 3  & 0.17  & 0.096 \\
Gd~\textsc{ii} & 64 & $-$0.30 & 0.25  & 3  & 0.14  & 0.081 &  & $-$0.35 & 0.35  & 3  & 0.16  & 0.092 &  & $-$0.41 & 0.36  & 3  & 0.11  & 0.061 \\
Tb~\textsc{ii} & 65 & $-$1.48 & $-$0.15 & 1  & 0.11  & 0.110 &  & $-$1.35 & 0.12  & 1  & 0.11  & 0.110 &  & $-$1.67 & $-$0.13 & 1  & 0.11  & 0.110 \\
Dy~\textsc{ii} & 66 & $-$0.20 & 0.33  & 4  & 0.13  & 0.063 &  & $-$0.19 & 0.48  & 4  & 0.09  & 0.046 &  & $-$0.37 & 0.37  & 4  & 0.29  & 0.144 \\
Ho~\textsc{ii} & 67 & $-$1.10 & 0.05  & 1  & 0.20  & 0.200 &  & $-$1.15 & 0.14  & 1  & 0.20  & 0.200 &  & $-$1.20 & 0.16  & 1  & 0.20  & 0.200 \\
Er~\textsc{ii} & 68 & $-$0.30 & 0.41  & 2  & 0.08  & 0.057 &  & $-$0.48 & 0.37  & 2  & 0.10  & 0.070 &  & $-$0.55 & 0.37  & 2  & 0.07  & 0.050 \\
Tm~\textsc{ii} & 69 & $-$1.43 & 0.10  & 2  & 0.18  & 0.125 &  & $-$1.53 & 0.15  & 2  & 0.25  & 0.175 &  & $-$1.65 & 0.09  & 2  & 0.07  & 0.050 \\
Yb~\textsc{ii} & 70 & $-$1.05 & $-$0.34 & 1  & 0.20  & 0.200 &  & $-$0.75 & 0.10  & 1  & 0.20  & 0.200 &  & $-$1.00 & $-$0.08 & 1  & 0.20  & 0.200 \\
Hf~\textsc{ii} & 72 & $-$1.00 & $-$0.22 & 1  & 0.11  & 0.110 &  & $-$0.93 & $-$0.01 & 1  & 0.11  & 0.110 &  & $-$1.09 & $-$0.10 & 1  & 0.11  & 0.110 \\
Ir~\textsc{i}  & 77 & 0.00  & 0.34  & 1  & 0.11  & 0.110 &  & $-$0.05 & 0.43  & 1  & 0.11  & 0.110 &  & 0.15  & 0.57  & 1  & 0.11  & 0.110 \\
Pb~\textsc{i}  & 82 & 0.05  & $-$0.27 & 1  & 0.11  & 0.110 &  & $-$0.01 & $-$0.19 & 1  & 0.11  & 0.110 &  & $-$0.08 & $-$0.32 & 1  & 0.11  & 0.110 \\
Th~\textsc{ii} & 90 & $-$1.55 & 0.02  & 1  & 0.11  & 0.110 &  & $-$1.42 & 0.29  & 1  & 0.11  & 0.110 &  & $-$1.46 & 0.32  & 1  & 0.11  & 0.110 \\
\enddata
\tablecomments{
Quoted uncertainties represent internal uncertainties only.
$\langle$[Fe/H]$\rangle$ is listed in the $\langle$[X/Fe]$\rangle$ column for
Fe~\textsc{i} and Fe~\textsc{ii}.
}
\end{deluxetable*}

Absolute uncertainties that account for errors in the derived atmospheric 
parameters are discussed in \citet{marino11} and presented in 
Table~4 of that work.
In the present study,
if only 1 line of a particular species has been measured
we adopt an uncertainty of 0.11~dex.
This estimate is based on the
mean standard deviation of individual lines for well-measured \ncap\ species
(i.e., $N \geq$~3).
Some lines in Table~\ref{lineabundtab} are marked with ``:'' to indicate that
the derived abundance is less certain due to significant blending
features, difficult continuum placement, etc.
These lines have an adopted internal uncertainty of 0.2~dex.

\begin{deluxetable*}{ccccccccccccccccccc}
\tablecaption{Mean Abundances in the Three $r+s$ Stars
\label{rsabundtab}
}
\tablewidth{0pt}
\tablehead{
\colhead{} &
\colhead{} &
\multicolumn{5}{c}{I-27} &
\colhead{} &
\multicolumn{5}{c}{I-53} &
\colhead{} &
\multicolumn{5}{c}{I-80} \\
\cline{3-7} \cline{9-13} \cline{15-19}
\colhead{Species} &
\colhead{$Z$} &
\colhead{$\langle\log\epsilon\rangle$} &
\colhead{$\langle$[X/Fe]$\rangle$} &
\colhead{$N$} &
\colhead{$\sigma$} &
\colhead{$\sigma_{\mu}$} &
\colhead{} &
\colhead{$\langle\log\epsilon\rangle$} &
\colhead{$\langle$[X/Fe]$\rangle$} &
\colhead{$N$} &
\colhead{$\sigma$} &
\colhead{$\sigma_{\mu}$} &
\colhead{} &
\colhead{$\langle\log\epsilon\rangle$} &
\colhead{$\langle$[X/Fe]$\rangle$} &
\colhead{$N$} &
\colhead{$\sigma$} &
\colhead{$\sigma_{\mu}$}
}
\startdata
Na~\textsc{i}  & 11 & 4.75  & 0.39  & 3   & 0.16  & 0.092 &  & 4.70  & 0.27  & 4  & 0.08  & 0.039 &  & 5.11  & 0.62  & 4  & 0.06  & 0.029 \\
Mg~\textsc{i}  & 12 & 6.13  & 0.41  & 3   & 0.26  & 0.150 &  & 6.28  & 0.49  & 3  & 0.15  & 0.085 &  & 6.25  & 0.40  & 3  & 0.18  & 0.105 \\
Al~\textsc{i}  & 13 & 5.02  & 0.45  & 4   & 0.23  & 0.115 &  & 5.09  & 0.45  & 4  & 0.13  & 0.063 &  & 5.57  & 0.87  & 4  & 0.10  & 0.048 \\
Si~\textsc{i}  & 14 & 6.02  & 0.39  & 3   & 0.09  & 0.053 &  & 5.88  & 0.18  & 3  & 0.21  & 0.123 &  & 5.94  & 0.18  & 3  & 0.19  & 0.109 \\
K~\textsc{i}   & 19 & 4.14  & 0.99  & 1   & 0.11  & 0.110 &  & 4.09  & 0.87  & 1  & 0.11  & 0.110 &  & 4.12  & 0.84  & 1  & 0.11  & 0.110 \\
Ca~\textsc{i}  & 20 & 5.00  & 0.54  & 8   & 0.10  & 0.035 &  & 4.99  & 0.46  & 8  & 0.08  & 0.028 &  & 5.04  & 0.45  & 8  & 0.11  & 0.038 \\
Sc~\textsc{ii} & 21 & 1.73  & 0.38  & 5   & 0.29  & 0.132 &  & 1.60  & 0.24  & 5  & 0.28  & 0.124 &  & 1.66  & 0.26  & 5  & 0.33  & 0.147 \\
Ti~\textsc{i}  & 22 & 3.05  & $-$0.02 & 9   & 0.20  & 0.066 &  & 3.26  & 0.12  & 9  & 0.14  & 0.048 &  & 3.26  & 0.06  & 9  & 0.12  & 0.040 \\
Ti~\textsc{ii} & 22 & 3.53  & 0.39  & 9   & 0.14  & 0.048 &  & 3.65  & 0.49  & 9  & 0.14  & 0.046 &  & 3.66  & 0.46  & 9  & 0.13  & 0.042 \\
V~\textsc{i}   & 23 & 2.06  & 0.02  & 5   & 0.15  & 0.066 &  & 1.98  & $-$0.14   & 5  & 0.11  & 0.051 &  & 2.02  & $-$0.16 & 5  & 0.16  & 0.072 \\
Cr~\textsc{i}  & 24 & 3.54  & $-$0.22 & 6   & 0.17  & 0.067 &  & 3.67  & $-$0.16 & 6  & 0.06  & 0.026 &  & 3.71  & $-$0.18 & 6  & 0.07  & 0.030 \\
Cr~\textsc{ii} & 24 & 3.99  & 0.15  & 1   & 0.11  & 0.110 &  & 4.05  & 0.20  & 1  & 0.11  & 0.110 &  & 4.26  & 0.37  & 1  & 0.11  & 0.110 \\
Mn~\textsc{i}  & 25 & 3.10  & $-$0.45 & 4   & 0.16  & 0.080 &  & 3.08  & $-$0.54 & 4  & 0.10  & 0.048 &  & 3.12  & $-$0.56 & 4  & 0.13  & 0.063 \\
Fe~\textsc{i}  & 26 & 5.62  & $-$1.88 & 69  & 0.17  & 0.020 &  & 5.68  & $-$1.82 & 69 & 0.12  & 0.015 &  & 5.75  & $-$1.75 & 69 & 0.13  & 0.016 \\
Fe~\textsc{ii} & 26 & 5.70  & $-$1.80 & 7   & 0.12  & 0.046 &  & 5.71  & $-$1.79 & 7  & 0.13  & 0.048 &  & 5.75  & $-$1.75 & 7  & 0.14  & 0.054 \\
Co~\textsc{i}  & 27 & 3.17  & 0.06  & 4   & 0.08  & 0.039 &  & 3.06  & $-$0.12 & 4  & 0.13  & 0.064 &  & 3.15  & $-$0.09 & 4  & 0.09  & 0.045 \\
Ni~\textsc{i}  & 28 & 4.24  & $-$0.10 & 10  & 0.10  & 0.032 &  & 4.23  & $-$0.17 & 10 & 0.08  & 0.027 &  & 4.31  & $-$0.16 & 10 & 0.08  & 0.025 \\
Cu~\textsc{i}  & 29 & 1.95  & $-$0.36 & 2   & 0.08  & 0.057 &  & 1.87  & $-$0.51 & 2  & 0.08  & 0.057 &  & 1.84  & $-$0.60 & 2  & 0.08  & 0.057 \\
Zn~\textsc{i}  & 30 & 2.89  & 0.21  & 2   & 0.08  & 0.057 &  & 3.03  & 0.28  & 2  & 0.11  & 0.080 &  & 3.08  & 0.27  & 2  & 0.10  & 0.070 \\
Rb~\textsc{i}  & 37 & $<$1.30 & $<$0.66 & 1   &\nodata&\nodata&  & $<$1.20 & $<$0.49 & 1  &\nodata&\nodata&  & $<$1.35 & $<$0.58 & 1  &\nodata&\nodata\\
Sr~\textsc{i}  & 38 & 1.08  & 0.09  & 1   & 0.11  & 0.110 &  & 1.08  & 0.02  & 1  & 0.11  & 0.110 &  & 0.97  & $-$0.15 & 1  & 0.11  & 0.110 \\
Y~\textsc{ii}  & 39 & 0.77  & 0.36  & 7   & 0.08  & 0.031 &  & 0.69  & 0.27  & 7  & 0.05  & 0.017 &  & 0.72  & 0.26  & 7  & 0.09  & 0.035 \\
Zr~\textsc{i}  & 40 & 1.07  & 0.37  & 2   & 0.08  & 0.057 &  & 1.18  & 0.40  & 2  & 0.11  & 0.075 &  & 1.10  & 0.27  & 2  & 0.08  & 0.057 \\
Zr~\textsc{ii} & 40 & 1.47  & 0.69  & 3   & 0.28  & 0.160 &  & 1.31  & 0.52  & 3  & 0.07  & 0.041 &  & 1.43  & 0.60  & 3  & 0.12  & 0.068 \\
Mo~\textsc{i}  & 42 & 0.32  & 0.32  & 3   & 0.45  & 0.259 &  & 0.37  & 0.30  & 3  & 0.55  & 0.388 &  & 0.18  & 0.05  & 2  & 0.23  & 0.130 \\
Ru~\textsc{i}  & 44 & 0.15  & 0.28  & 2   & 0.08  & 0.057 &  & 0.00  & 0.06  & 1  & 0.14  & 0.100 &  & 0.50  & 0.50  & 1  & 0.14  & 0.100 \\
Rh~\textsc{i}  & 45 & $-$0.45 & 0.37  & 1   & 0.11  & 0.110 &  & $-$0.55 & 0.20  & 1  & 0.20  & 0.200 &  & $-$0.80 & $-$0.11 & 1  & 0.11  & 0.110 \\
Ba~\textsc{ii} & 56 & 1.20  & 0.82  & 2   & 0.08  & 0.057 &  & 1.13  & 0.74  & 2  & 0.08  & 0.057 &  & 1.09  & 0.66  & 2  & 0.08  & 0.057 \\
La~\textsc{ii} & 57 & $-$0.11 & 0.59  & 14  & 0.14  & 0.038 &  & $-$0.13 & 0.56  & 14 & 0.07  & 0.020 &  & $-$0.21 & 0.44  & 14 & 0.08  & 0.021 \\
Ce~\textsc{ii} & 58 & 0.25  & 0.48  & 16  & 0.13  & 0.032 &  & 0.43  & 0.64  & 16 & 0.12  & 0.029 &  & 0.21  & 0.38  & 16 & 0.11  & 0.028 \\
Pr~\textsc{ii} & 59 & $-$0.60 & 0.49  & 4   & 0.06  & 0.029 &  & $-$0.61 & 0.47  & 4  & 0.06  & 0.030 &  & $-$0.72 & 0.31  & 4  & 0.06  & 0.030 \\
Nd~\textsc{ii} & 60 & 0.22  & 0.61  & 24  & 0.10  & 0.020 &  & 0.19  & 0.56  & 24 & 0.09  & 0.019 &  & 0.06  & 0.39  & 24 & 0.09  & 0.019 \\
Sm~\textsc{ii} & 62 & $-$0.38 & 0.46  & 9   & 0.12  & 0.039 &  & $-$0.33 & 0.50  & 9  & 0.09  & 0.029 &  & $-$0.38 & 0.41  & 9  & 0.09  & 0.029 \\
Eu~\textsc{ii} & 63 & $-$1.05 & 0.23  & 3   & 0.26  & 0.149 &  & $-$0.91 & 0.36  & 3  & 0.13  & 0.073 &  & $-$0.91 & 0.32  & 3  & 0.11  & 0.065 \\
Gd~\textsc{ii} & 64 & $-$0.25 & 0.48  & 3   & 0.11  & 0.065 &  & $-$0.15 & 0.57  & 3  & 0.12  & 0.070 &  & $-$0.32 & 0.37  & 2  & 0.22  & 0.155 \\
Tb~\textsc{ii} & 65 &\nodata&\nodata& 0   &\nodata&\nodata&  & $-$1.28 & 0.21  & 1  & 0.11  & 0.110 &  & $-$1.30 & 0.15  & 1  & 0.11  & 0.110 \\
Dy~\textsc{ii} & 66 & $-$0.25 & 0.46  & 3   & 0.14  & 0.079 &  & $-$0.08 & 0.61  & 4  & 0.13  & 0.067 &  & $-$0.16 & 0.49  & 3  & 0.06  & 0.035 \\
Ho~\textsc{ii} & 67 & $-$1.15 & 0.17  & 1   & 0.20  & 0.200 &  & $-$1.18 & 0.13  & 1  & 0.20  & 0.200 &  & $-$1.20 & 0.07  & 1  & 0.20  & 0.200 \\
Er~\textsc{ii} & 68 & $-$0.41 & 0.47  & 2   & 0.16  & 0.110 &  & $-$0.43 & 0.45  & 2  & 0.19  & 0.135 &  & $-$0.25 & 0.59  & 2  & 0.13  & 0.095 \\
Tm~\textsc{ii} & 69 &\nodata&\nodata& 0   &\nodata&\nodata&  & $-$1.50 & 0.19  & 1  & 0.20  & 0.200 &  &\nodata&\nodata& 0  &\nodata&\nodata\\
Yb~\textsc{ii} & 70 & $-$0.76 & 0.12  & 1   & 0.11  & 0.110 &  & $-$0.56 & 0.31  & 1  & 0.11  & 0.110 &  & $-$0.52 & 0.31  & 1  & 0.11  & 0.110 \\
Hf~\textsc{ii} & 72 & $-$0.56 & 0.39  & 1   & 0.11  & 0.110 &  & $-$0.55 & 0.39  & 1  & 0.11  & 0.110 &  & $-$0.60 & 0.30  & 1  & 0.11  & 0.110 \\
Ir~\textsc{i}  & 77 &\nodata&\nodata& 0   &\nodata&\nodata&  &\nodata&\nodata& 0  &\nodata&\nodata&  &\nodata&\nodata& 0  &\nodata&\nodata\\
Pb~\textsc{i}  & 82 & 0.77  & 0.61  & 1   & 0.11  & 0.110 &  & 0.97  & 0.74  & 1  & 0.11  & 0.110 &  & 0.63  & 0.34  & 1  & 0.11  & 0.110 \\
Th~\textsc{ii} & 90 &\nodata&\nodata& 0   &\nodata&\nodata&  &\nodata&\nodata& 0  &\nodata&\nodata&  &\nodata&\nodata& 0  &\nodata&\nodata\\
\enddata
\tablecomments{
Quoted uncertainties represent internal uncertainties only.  
$\langle$[Fe/H]$\rangle$ is listed in the $\langle$[X/Fe]$\rangle$ column for
Fe~\textsc{i} and Fe~\textsc{ii}.
}
\end{deluxetable*}

One difficulty that is not minimized by our approach is 
that of comparing abundance ratios derived from species of different
ionization states.
Ratios of, e.g., [Ca/Fe] or [Eu/Fe] are computed by comparing 
Ca~\textsc{i} to Fe~\textsc{i} or
Eu~\textsc{ii} to Fe~\textsc{ii}
since both species of Fe are detected. 
Other ratios, such as [Pb/La] or [Pb/Eu], which compare 
Pb~\textsc{i} to La~\textsc{ii} or Eu~\textsc{ii},
may be systematically uncertain.
Note that \textit{for illustration purposes in the figures only}
we normalize the abundances of first-peak \ncap\ elements
observed in their neutral state
(Sr~\textsc{i}, Mo~\textsc{i}, 
Ru~\textsc{i}, and Rh~\textsc{i}) 
to the singly-ionized abundances by
the difference in Zr~\textsc{ii} and Zr~\textsc{i}
in each star (typically 0.2--0.4~dex).

\section{Heavy Element Abundances in M22}
\label{results}

In this section we analyze the abundance patterns in detail.
The abundance results for each star in the $r$-only and $r+s$ groups are
given in Tables~\ref{rabundtab} and \ref{rsabundtab}, respectively.
Table~\ref{meantab} lists the mean abundances for each element
in the $r$-only and $r+s$ groups.
We derive only upper limits from the Rb~\textsc{i} line at 7800\AA.
Due to blending by CN and CH, we are unable to derive
abundances of Ir~\textsc{i} or Th~\textsc{ii} in 
any star in the $r+s$ group.

\subsection{The Light and Fe-group Abundance Patterns in M22}
\label{fegroup}

\begin{figure}
\begin{center}
\includegraphics[angle=0,width=3.2in]{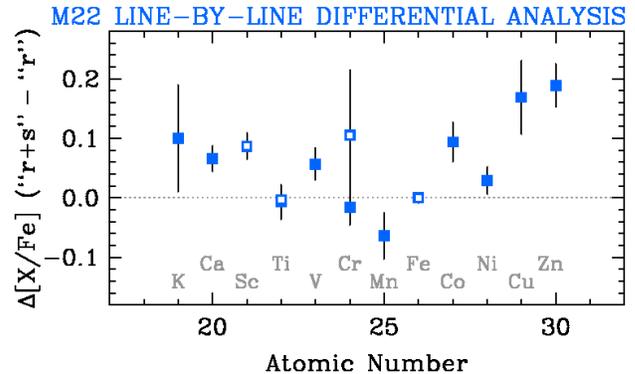}
\end{center}
\caption{
\label{lightdiffplot}
Differences in the mean abundances between the 3 $r$-only stars
and the 3 $r+s$ stars as a function of atomic number
for K through Zn.
Solid squares indicate neutral species, and
open squares indicate singly-ionized species.
The dotted line indicates zero difference.
}
\end{figure}

We derive abundances of Na~\textsc{i}, Mg~\textsc{i}, 
Al~\textsc{i}, and Si~\textsc{i} in each star of the sample.
After accounting for differences in the $\log(gf)$ values between
this study and \citet{marino11}, these abundances are 
in agreement within the uncertainties.  
\citeauthor{marino09}\ have discussed these abundances
at length, so we shall not consider them further.

\citet{marino11} detected an enhancement by 0.10~dex in the 
[Ca/Fe] ratio in the $r+s$ group of stars in M22, which we recover
in our data.
Other neighboring elements not included in that study
also exhibit very slight differences in our data.
To quantify these differences, we apply a line-by-line differential analysis,
which is largely insensitive to uncertainties in the
$\log(gf)$ values and star-to-star systematic effects
in the abundance analysis. 
The differential results are listed in Table~\ref{fegrouptab}
and illustrated in Figure~\ref{lightdiffplot}.
When considering the standard error ($\sigma_{\mu} \equiv \sigma/\sqrt{N}$)
of the mean line-by-line differential abundances (column~5), 
[K/Fe], [Ca/Fe], [Sc/Fe],\footnote{
The Sc~\textsc{ii} lines give discordant abundances, 
which may indicate relatively large 
uncertainties in the $\log(gf)$ values, but
the line-by-line results are extremely consistent.}
and [V/Fe] show slight
but significant (0.06--0.10~dex) enhancements in the $r+s$ group,
[Ti/Fe] and [Cr/Fe] are indistinguishable in the two groups,
and [Mn/Fe] shows a slight (0.06~dex) deficiency in the $r+s$ group.

\citet{marino11} saw an increase by 0.06--0.15~dex
in the [Cu/Fe] and [Zn/Fe] ratios of the $r+s$ group, which we also detect.
Furthermore, we find a similar---though smaller---enhancement
in [Co/Fe] and [Ni/Fe].
The results from a line-by-line differential
analysis of these elements are also listed in Table~\ref{fegrouptab}
and shown in Figure~\ref{lightdiffplot}.

A slight abundance enhancement in the Fe-group elements heavier than 
Fe may not be surprising, since \spro\ nucleosynthesis produces
heavy nuclei from successive neutron capture on Fe-group seeds.
Variations in the lighter Fe-group elements are more surprising.
We return to this issue in Section~\ref{agbmass}.

\subsection{The Neutron-Capture Abundance Patterns in M22}
\label{ncap}

\begin{figure*}
\begin{center}
\includegraphics[angle=0,width=2.8in]{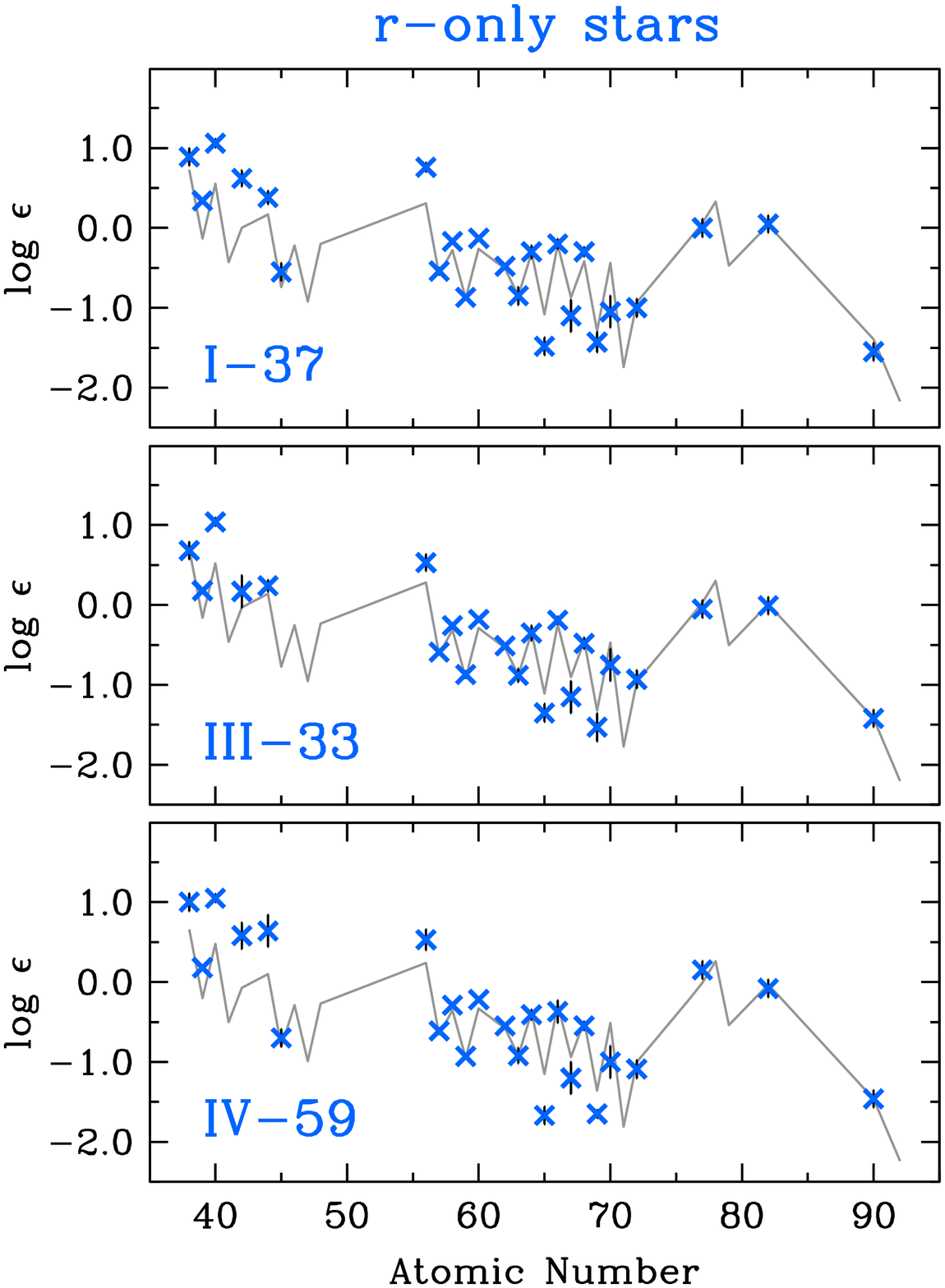} 
\hspace*{0.1in}
\includegraphics[angle=0,width=2.8in]{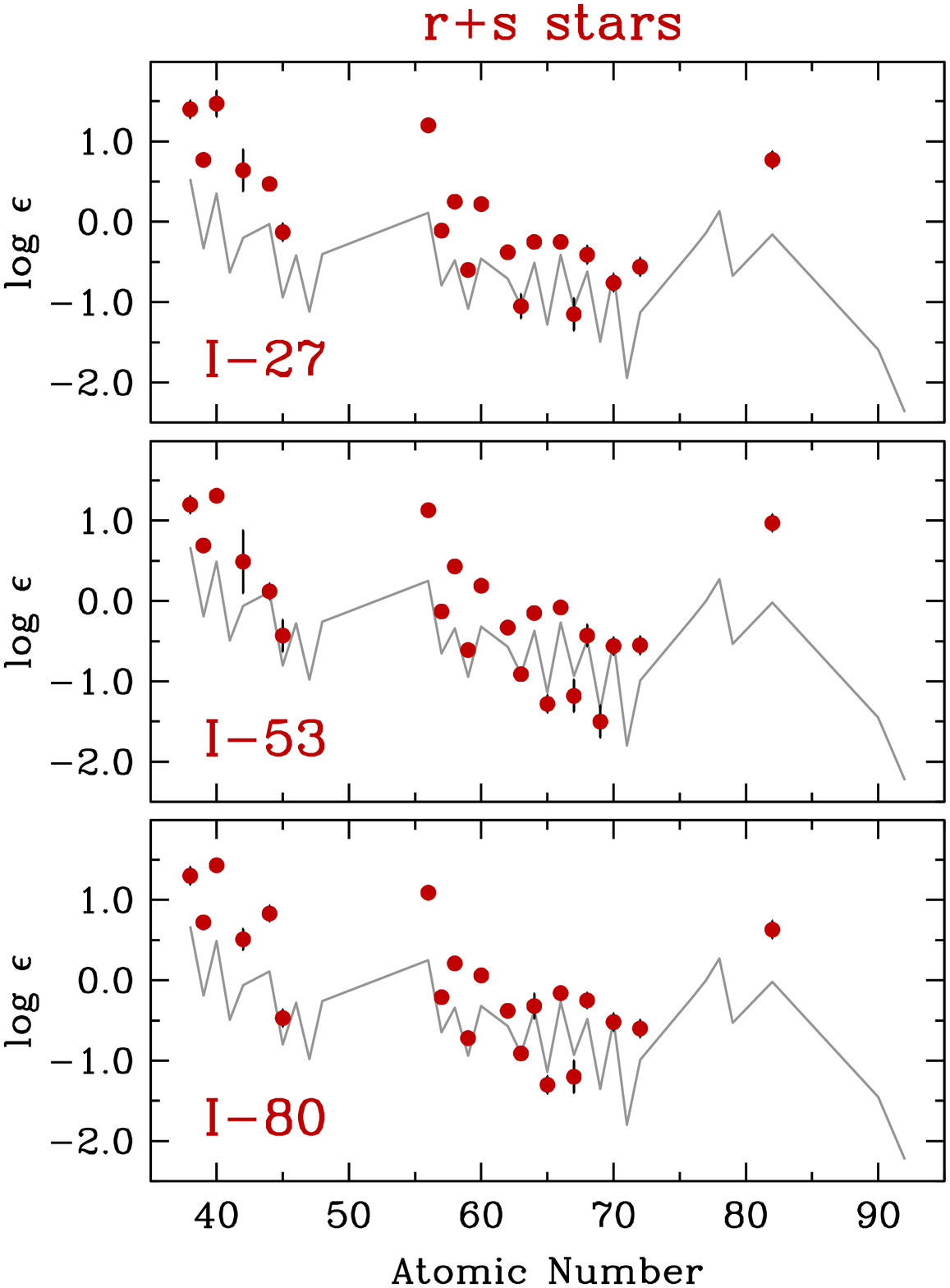}
\end{center}
\caption{
\label{starplot}
Logarithmic abundances for $Z \geq$~38 elements in the 
3 $r$-only stars (blue crosses, left panels) and the
3 $r+s$ stars (red circles, right panels)
as a function of atomic number.
The gray line illustrates the abundances in the \rpro\ standard star \bd\
\citep{cowan02,cowan05,sneden09,roederer10c}.
Pb has not been detected in \bd, so we instead show the 
predicted Pb/Eu ratio based on the average Pb/Eu observed
in Figure~3 of \citet{roederer10b}.
The \bd\ abundance pattern has been normalized to the Eu abundance
in each star.
}
\end{figure*}
 
\begin{figure*}
\begin{center}
\includegraphics[angle=0,width=5.0in]{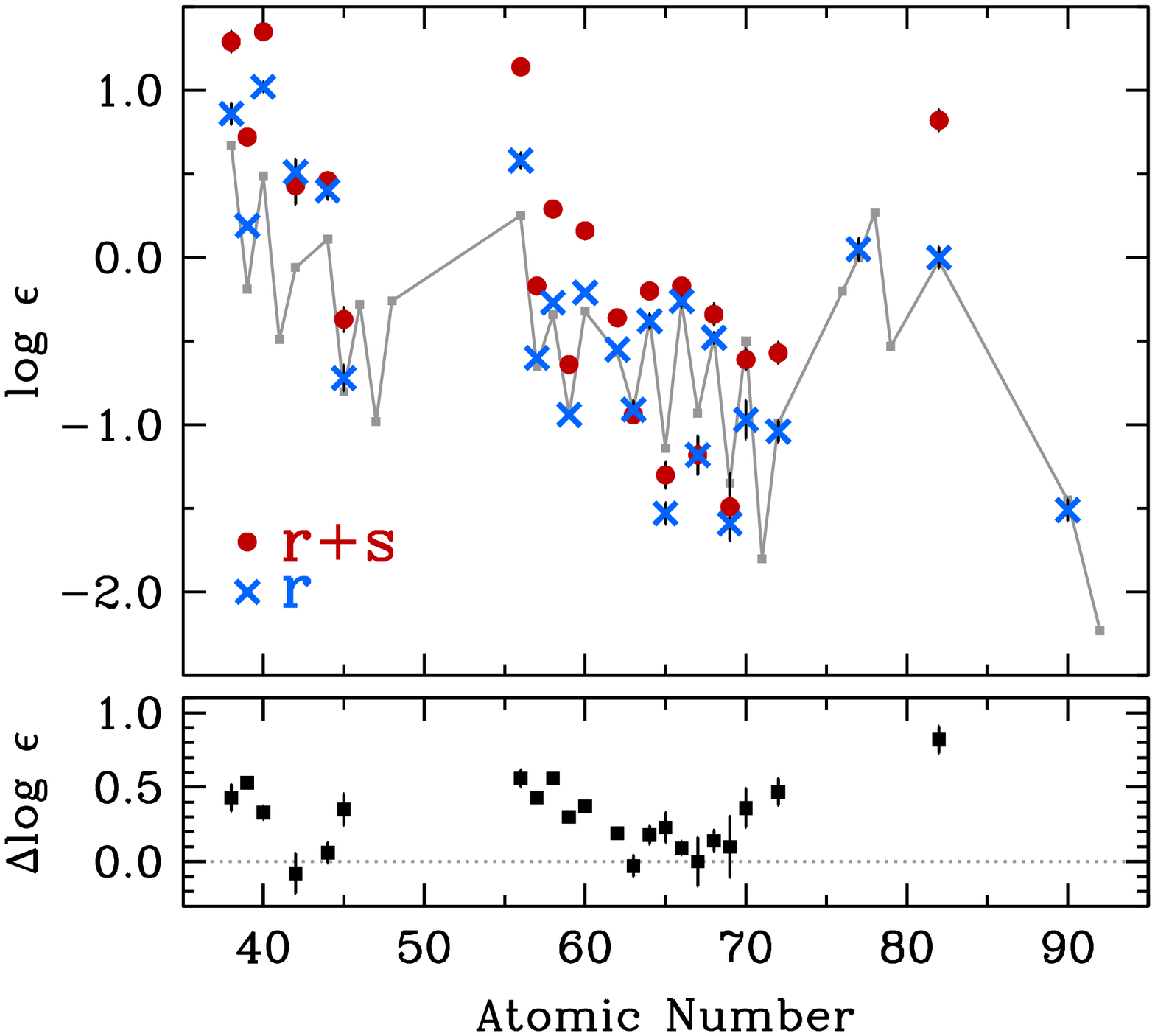}
\end{center}
\caption{
\label{meanplot}
Top panel: The mean logarithmic abundances for the 3 $r$-only stars
(blue crosses) and the 3 $r+s$ stars (red circles)
as a function of atomic number.
The gray line and small gray squares illustrate 
the abundances in the \rpro\ standard star \bd\
\citep{cowan02,cowan05,sneden09,roederer10c}.
Pb has not been detected in \bd, so we instead show the 
predicted Pb/Eu ratio based on the average Pb/Eu observed
in Figure~3 of \citet{roederer10b}.
The \bd\ abundance pattern has been normalized to the Eu abundance.
Bottom panel: The differences in these mean abundances.
The dotted line indicates zero difference.
}
\end{figure*}
 
Figure~\ref{starplot} illustrates the abundance patterns for the 
$Z \geq$~38 elements in each of the 6~stars observed in M22. 
The abundance pattern of the metal-poor
([Fe/H]~$= -$2.1) \rpro-rich standard star \bd\ 
([Eu/Fe]~$= +$0.9) is shown for comparison.
The three stars selected from the $s$-poor group of \citet{marino11}---our
``$r$-only'' group---share a similar abundance pattern with each other
and \bd.
The three stars from the \citeauthor{marino11}\ $s$-rich group---our
``$r+s$ group''---share a similar abundance pattern with each other
that clearly differs from \bd\ for the lighter \ncap\ elements and Pb.
Figure~\ref{starplot} demonstrates that it is appropriate to average
together the abundances of the 3~stars in each of these two groups
to reduce random uncertainties in the abundances, particularly for
the abundances derived from small numbers of lines.
The average abundance patterns for the $r$-only and $r+s$ groups are 
shown in Figure~\ref{meanplot}.
The derived mean [Fe/H] for the 3~stars in each group is the same, 
so the relative vertical scaling
of the abundances in Figure~\ref{meanplot} is not affected by the
bulk metal content of these two groups.

In the $r$-only group, 
the abundance pattern for Ba and the heavier elements ($Z \geq$~56)
generally conforms to that of \bd.
When normalized to Eu ($Z =$~63), 
the Ba, Ce, and Nd ($Z =$~56, 58, and 60, respectively)
abundances in the $r$-only group appear slightly
enhanced relative to \bd.
Furthermore, in the $r$-only group
several of the odd-$Z$ elements in the rare earth domain
(Tb, Ho, and Tm---elements 65, 67, and 69, respectively)
plus the even-$Z$ element Yb ($Z =$~70) lie 0.2--0.4~dex
below the \bd\ abundances.
This is not surprising given that the \rpro\ enrichment in M22
is less extreme than that seen in \bd\ or other $r$-rich standard stars,
and variations in the physical conditions at the time of the
nucleosynthesis may be responsible \citep{roederer10b}.
The abundances of the lighter elements Sr--Rh (38~$\leq Z \leq$~45) are known
to vary widely among metal-poor stars that show no evidence 
of \spro\ enrichment (e.g., \citealt{roederer10b} and references therein).
Based on the empirical correlation between [Eu/Y] and [Eu/Fe]
identified by \citet{barklem05}, \citet{otsuki06}, \citet{montes07}, and
\citet{roederer10b}, we would expect the Sr--Rh elements in the M22
$r$-only group to be more abundant than those in \bd\
when normalized to Eu, which is indeed the case.
These elements may be produced by primary nucleosynthetic 
mechanisms in addition to the \rpro\ 
(e.g., charged-particle reactions in the expanding neutrino winds of
core collapse SNe; \citealt{woosley92})
and so could be expected to vary.

In the $r+s$ group,
all heavy elements except Mo ($Z =$~42), Ru ($Z =$~44), 
Eu, Ho, and Tm are enhanced relative to the $r$-only group.
These differences are most pronounced among the lightest
\ncap\ elements (Sr, Y, and Zr), the 
light and heavy ends of the rare earth domain 
(Ba--Nd and Yb--Hf), and Pb.
This is not surprising, given that a significant fraction of the S.S.\
abundance of each of these elements is attributed to
the \spro.
In contrast, the S.S.\ abundances of
elements in the middle of the rare earth domain
are mostly attributed to the \rpro.

\begin{figure*}
\begin{center}
\includegraphics[angle=270,width=4.2in]{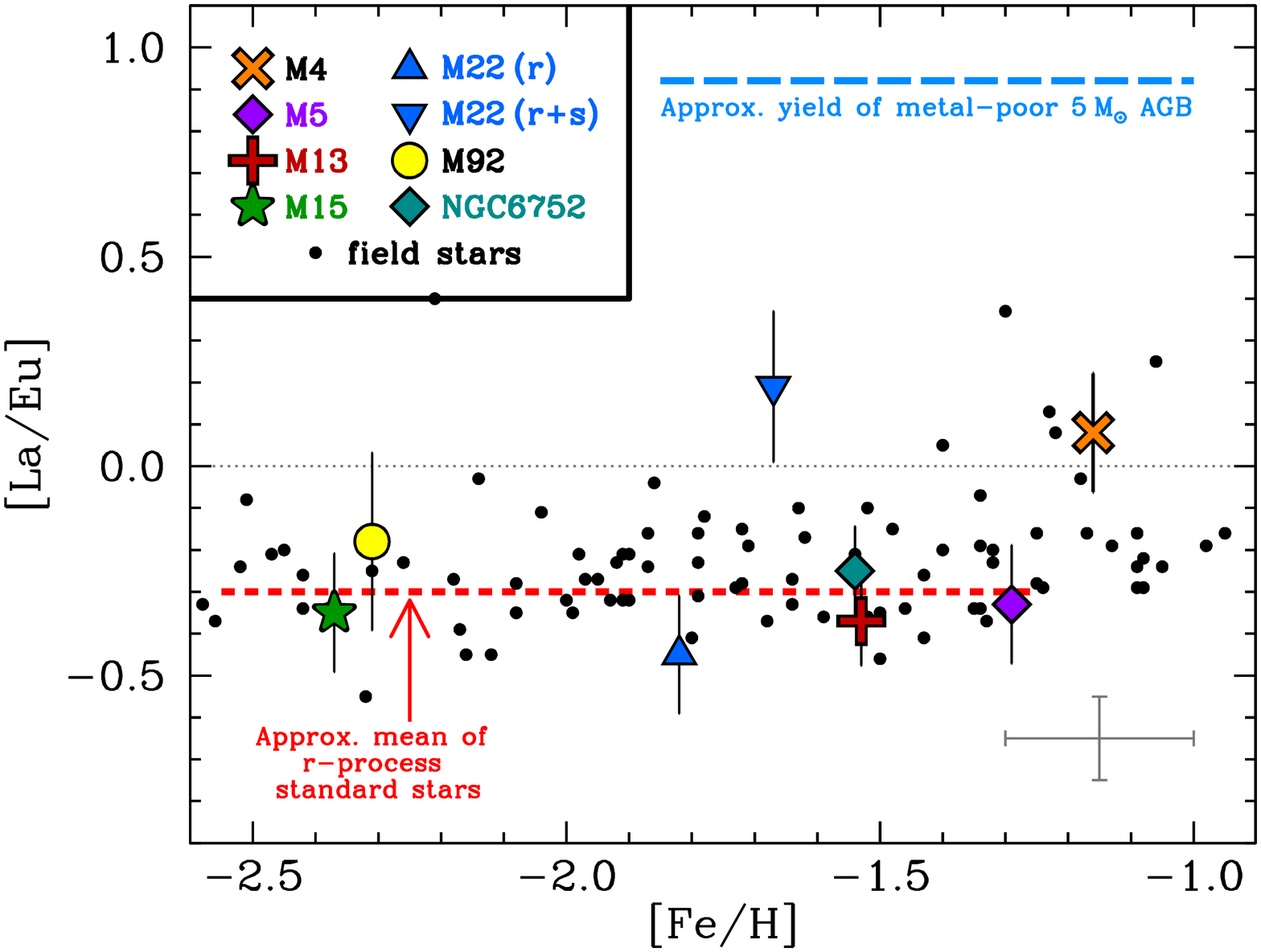}  \\
\vspace*{0.1in}
\includegraphics[angle=270,width=4.2in]{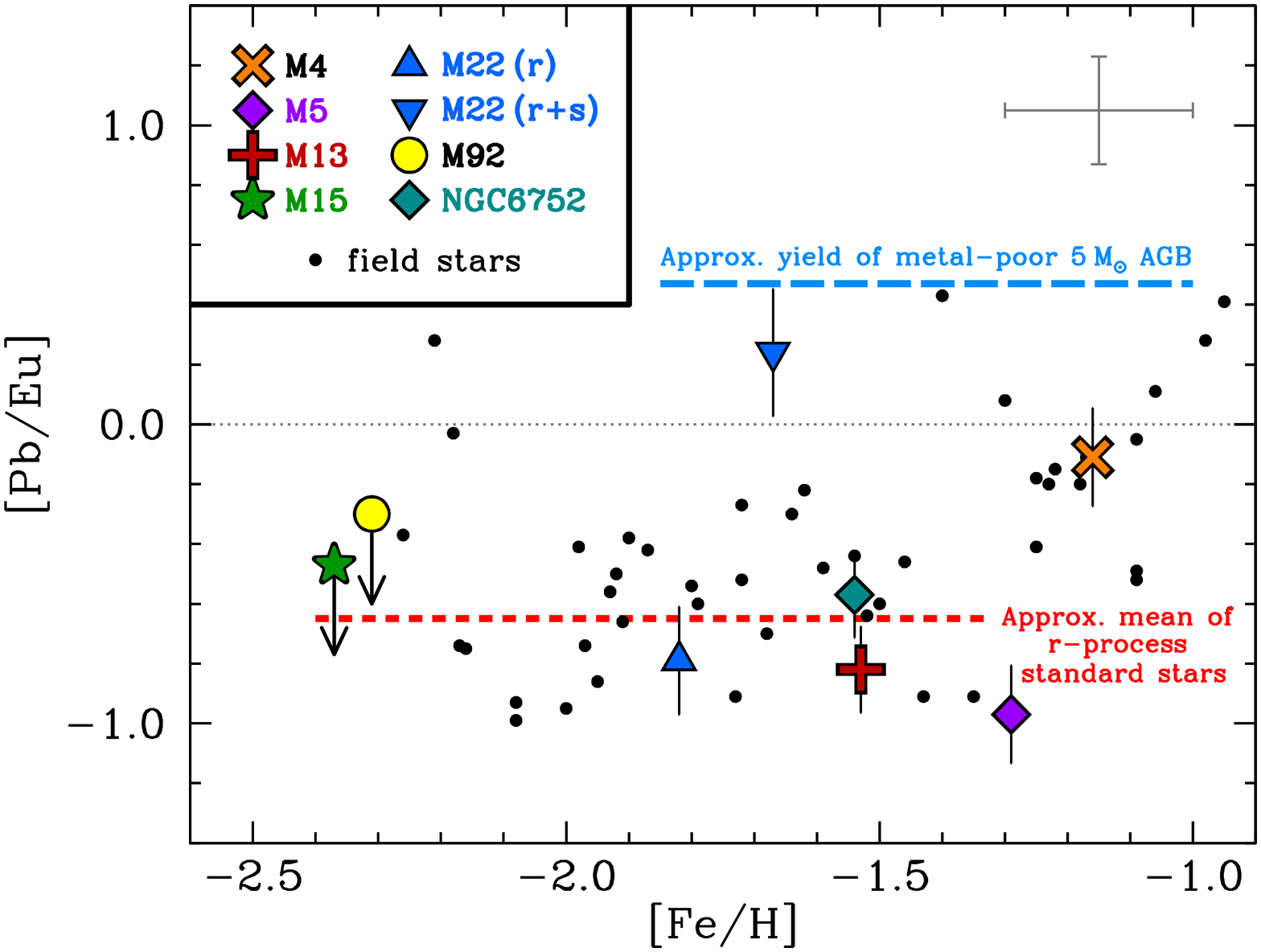}
\end{center}
\caption{
\label{gcpbplot}
[La/Eu] and [Pb/Eu] ratios as a function of [Fe/H].
Only GCs where Pb has been measured have been included.
The dotted lines indicate the solar ratio.
A typical uncertainty is shown.
The blue long-dashed lines indicate the approximate yields of a 5~\msun\ 
AGB star at [Fe/H]~$= -$2.3 \citep{roederer10b}.
The red short-dashed lines indicate the approximate means of metal-poor
field stars whose [Pb/Eu] ratios are consistent with
having been enriched by \rpro\ material only \citep{roederer10b}.
GC abundances are referenced as follows:
M4 and M5, \citet{yong08a,yong08b};
M13 and \mbox{NGC~6752}, \citet{yong06};
M15, \citet{sobeck11};
M92, \citet{roederer11b};
M22, this study;
field stars, \citet{roederer10b}.
All abundances have been normalized to the scale used in the present study.
}
\end{figure*}

Low-metallicity AGBs produce substantial overabundances
of Pb relative to the Fe-group \spro\ seeds and all
elements intermediate between Fe and Pb
(e.g., \citealt{clayton88,gallino98}).
As the metallicity of the \spro\ environment increases
above [Fe/H]~$\sim -$1.0, the Pb overabundances decrease
\citep{travaglio01}.
\citet{roederer10b} have shown that [Pb/Eu] ratios can be
an effective diagnostic to 
identify low metallicity stars that lack
detectable contributions from the \spro.
It is clear from Figures~\ref{pbplot}, \ref{starplot}, and \ref{meanplot}
that the Pb abundance is moderately
enhanced in the $r+s$ group of stars relative to the $r$-only group.
As shown in Figure~\ref{gcpbplot},
[La/Eu] and [Pb/Eu] in M5, M13, M15, M92, and \mbox{NGC~6752}
\citep{yong06,yong08a,yong08b,sobeck11,roederer11b}
are the same as that for field
stars of the same metallicity.
These ratios are low and
suggest no contribution from \spro\ material.
The M22 $r$-only group is normal for other metal-poor
GCs in this regard.
[Pb/Eu] is moderately enhanced in the M22 $r+s$ group,
and this increase relative to the $r$-only
group (a difference of $+$0.85~dex) 
is notably higher than other [X/Eu] ratios
($\leq +$0.55~dex).
This further confirms the results of \citet{marino09,marino11}
that the $r+s$ (or $s$-rich) group in M22 
contains a moderate amount of \spro\ material.

\subsection{The Age of M22 Calculated from Radioactive $^{232}$Th Decay}
\label{age}

The radioactive isotope $^{232}$Th can only
be produced in \rpro\ nucleosynthesis.
It can be used in conjunction with other stable 
elements produced in the same events
to yield an age for the \rpro\ material in M22.
This can be done in a relative sense (e.g., comparing the
Th/Eu ratio in several GCs) or an absolute sense if the
initial production ratio of Th/Eu is known from theory.
We use the production ratio predicted by the
simulations of \citet{kratz07} and the derived $\log\epsilon$(Th/Eu) ratio 
in the 3~$r$-only stars in M22
($-$0.60~$\pm$~0.085) to calculate
an absolute age of 12.4~$\pm$~4.0~Gyr. 
Recall we could not measure Th in the $r+s$ group due to blending features.
This assumes no uncertainty in the initial production ratio, 
which likely
translates to an uncertainty of several Gyr
(e.g.,
\citealt{frebel07,kratz07,ludwig10}).
This age estimate
is consistent with the relatively old age derived from
isochrone fitting to the M22 main sequence turnoff \citep{marinfranch09},
the ages of other metal-poor GCs derived from their Th/Eu ratios
(\citealt{sneden00}, \citealt{johnson01}, \citealt{yong08b}, 
\citealt{lai11}),
and halo field stars of similar low metallicity (e.g., \citealt{roederer09}).
While the usefulness of this measurement is limited by
observational uncertainties and systematic effects,
the general agreement is reassuring.

 \begin{deluxetable}{ccccccc}
\tablecaption{Mean Abundances in the $r$ and $r+s$ Groups
\label{meantab}
}
\tablewidth{0pt}
\tablehead{
\colhead{} &
\colhead{} &
\multicolumn{2}{c}{$r$-only} &
\colhead{} &
\multicolumn{2}{c}{$r+s$} \\
\cline{3-4} \cline{6-7}
\colhead{Species} &
\colhead{$Z$} &
\colhead{$\langle$[X/Fe]$\rangle$} &
\colhead{$\sigma_{\mu}$} &
\colhead{} &
\colhead{$\langle$[X/Fe]$\rangle$} &
\colhead{$\sigma_{\mu}$} }
\startdata
Na~\textsc{i}  & 11 & 0.08  & 0.018  & &  0.49   & 0.023   \\
Mg~\textsc{i}  & 12 & 0.39  & 0.064  & &  0.45   & 0.060   \\
Al~\textsc{i}  & 13 & 0.34  & 0.034  & &  0.69   & 0.036   \\
Si~\textsc{i}  & 14 & 0.14  & 0.064  & &  0.33   & 0.044   \\
K~\textsc{i}   & 19 & 0.80  & 0.064  & &  0.90   & 0.064   \\
Ca~\textsc{i}  & 20 & 0.40  & 0.023  & &  0.48   & 0.019   \\
Sc~\textsc{ii} & 21 & 0.21  & 0.073  & &  0.30   & 0.077   \\
Ti~\textsc{i}  & 22 & 0.06  & 0.015  & &  0.07   & 0.028   \\
Ti~\textsc{ii} & 22 & 0.47  & 0.020  & &  0.45   & 0.026   \\
V~\textsc{i}   & 23 & $-$0.14 & 0.026  & &  $-$0.10  & 0.035   \\
Cr~\textsc{i}  & 24 & $-$0.14 & 0.020  & &  $-$0.17  & 0.019   \\
Cr~\textsc{ii} & 24 & 0.14  & 0.064  & &  0.24   & 0.064   \\
Mn~\textsc{i}  & 25 & $-$0.50 & 0.034  & &  $-$0.53  & 0.034   \\
Fe~\textsc{i}  & 26 & $-$1.81 & 0.008  & &  $-$1.81  & 0.009   \\
Fe~\textsc{ii} & 26 & $-$1.78 & 0.022  & &  $-$1.78  & 0.028   \\
Co~\textsc{i}  & 27 & $-$0.15 & 0.033  & &  $-$0.03  & 0.027   \\
Ni~\textsc{i}  & 28 & $-$0.20 & 0.020  & &  $-$0.15  & 0.016   \\
Cu~\textsc{i}  & 29 & $-$0.66 & 0.033  & &  $-$0.49  & 0.033   \\
Zn~\textsc{i}  & 30 & 0.04  & 0.032  & &  0.24   & 0.039   \\
Rb~\textsc{i}  & 37 &$<$0.20  & \nodata& & $<$0.49   & \nodata \\     
Sr~\textsc{i}  & 38 & $-$0.53 & 0.064  & &  $-$0.01  & 0.064   \\
Y~\textsc{ii}  & 39 & $-$0.24 & 0.013  & &  0.29   & 0.014   \\
Zr~\textsc{i}  & 40 & $-$0.08 & 0.036  & &  0.34   & 0.035   \\
Zr~\textsc{ii} & 40 & 0.22  & 0.030  & &  0.55   & 0.034   \\
Mo~\textsc{i}  & 42 & 0.11  & 0.078  & &  0.12   & 0.111   \\
Ru~\textsc{i}  & 44 & 0.13  & 0.053  & &  0.28   & 0.044   \\
Rh~\textsc{i}  & 45 & $-$0.30 & 0.078  & &  0.14   & 0.072   \\
Ba~\textsc{ii} & 56 & 0.18  & 0.047  & &  0.74   & 0.033   \\
La~\textsc{ii} & 57 & 0.08  & 0.016  & &  0.51   & 0.013   \\
Ce~\textsc{ii} & 58 & $-$0.07 & 0.009  & &  0.49   & 0.017   \\
Pr~\textsc{ii} & 59 & 0.12  & 0.016  & &  0.42   & 0.017   \\
Nd~\textsc{ii} & 60 & 0.15  & 0.009  & &  0.52   & 0.011   \\
Sm~\textsc{ii} & 62 & 0.27  & 0.013  & &  0.46   & 0.018   \\
Eu~\textsc{ii} & 63 & 0.35  & 0.056  & &  0.32   & 0.046   \\
Gd~\textsc{ii} & 64 & 0.33  & 0.043  & &  0.51   & 0.046   \\
Tb~\textsc{ii} & 65 & $-$0.05 & 0.064  & &  0.18   & 0.078   \\
Dy~\textsc{ii} & 66 & 0.42  & 0.036  & &  0.51   & 0.029   \\
Ho~\textsc{ii} & 67 & 0.12  & 0.115  & &  0.12   & 0.115   \\
Er~\textsc{ii} & 68 & 0.38  & 0.033  & &  0.52   & 0.063   \\
Tm~\textsc{ii} & 69 & 0.09  & 0.045  & &  0.19   & 0.200   \\
Yb~\textsc{ii} & 70 & $-$0.11 & 0.115  & &  0.25   & 0.064   \\
Hf~\textsc{ii} & 72 & $-$0.11 & 0.064  & &  0.36   & 0.064   \\
Ir~\textsc{i}  & 77 & 0.45  & 0.064  & &  \nodata& \nodata \\     
Pb~\textsc{i}  & 82 & $-$0.26 & 0.064  & &  0.56   & 0.064   \\
Th~\textsc{ii} & 90 & 0.21  & 0.064  & &  \nodata& \nodata \\     
\enddata
\tablecomments{
The means represent weighted means from the 3~stars in each 
group, and the stated uncertainties represent internal uncertainties only.
$\langle$[Fe/H]$\rangle$ is listed in the $\langle$[X/Fe]$\rangle$ column for
Fe~\textsc{i} and Fe~\textsc{ii}.
}
\end{deluxetable}

\section{Comparison to Other Complex Metal-poor GCs}
\label{merger}

\begin{figure}
\begin{center}
\includegraphics[angle=0,width=3.0in]{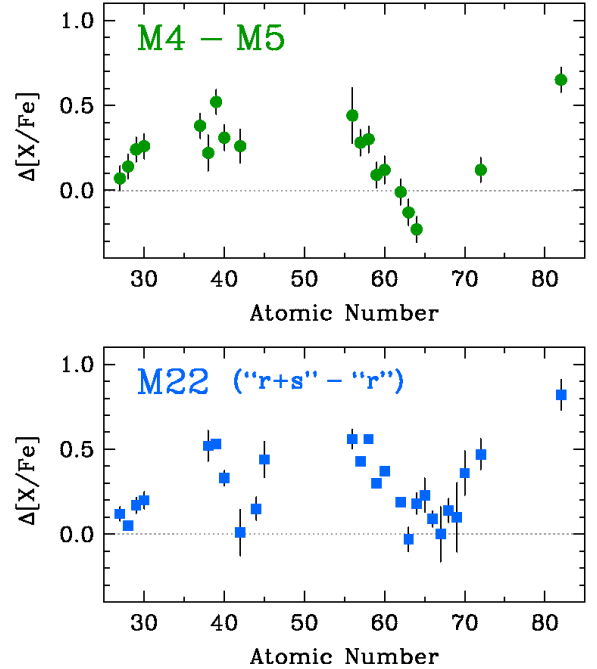}
\end{center}
\caption{
\label{m4m5plot}
Top panel: Differences between the mean abundances in GCs M4 and M5.
The abundances are taken from \citet{yong08a,yong08b} and
\citet{ivans99,ivans01}.
Bottom panel: Differences between the mean abundances in the
$r+s$ and $r$ groups in M22.
In both panels, dotted lines indicate zero difference.
}
\end{figure}

\begin{deluxetable}{lcccc}
\tablecaption{Mean Line-by-line Differentials for K--Zn in the $r$ and $r+s$ Groups
\label{fegrouptab}
}
\tablewidth{0pt}
\tablehead{
\colhead{Species} &
\colhead{$N$} &
\colhead{$\langle\Delta$[X/Fe]$\rangle$\tablenotemark{a}} &
\colhead{$\sigma$} &
\colhead{$\sigma_{\mu}$} }
\startdata
K~\textsc{i}      & 1 & $+$0.100 & 0.090 & 0.090 \\
Ca~\textsc{i}     & 8 & $+$0.066 & 0.063 & 0.022 \\
Sc~\textsc{ii}    & 5 & $+$0.087 & 0.048 & 0.022 \\
Ti~\textsc{i}     & 9 & $-$0.007 & 0.087 & 0.029 \\
Ti~\textsc{ii}    & 9 & $-$0.003 & 0.062 & 0.021 \\
Ti~\textsc{i$+$ii}&18 & $-$0.005 & 0.073 & 0.017 \\
V~\textsc{i}      & 5 & $+$0.057 & 0.060 & 0.027 \\
Cr~\textsc{i}     & 6 & $-$0.016 & 0.073 & 0.030 \\
Cr~\textsc{ii}    & 1 & $+$0.105 & 0.110 & 0.110 \\
Cr~\textsc{i$+$ii}& 7 & $-$0.008 & 0.085 & 0.032 \\
Mn~\textsc{i}     & 4 & $-$0.064 & 0.078 & 0.039 \\
Co~\textsc{i}     & 4 & $+$0.094 & 0.065 & 0.033 \\
Ni~\textsc{i}     &10 & $+$0.029 & 0.073 & 0.023 \\
Cu~\textsc{i}     & 2 & $+$0.169 & 0.088 & 0.062 \\
Zn~\textsc{i}     & 2 & $+$0.189 & 0.052 & 0.036 \\
\enddata
\tablenotetext{a}{
In the sense of $\langle$[X/Fe]$_{r+s}\rangle - \langle$[X/Fe]$_{r}\rangle$
}
\end{deluxetable}

As discussed in \citet{marino09,marino11}, evidence for
multiple stellar populations in M22 includes the following:
(1) the SGB shows two distinct sequences,
(2) there is a metallicity offset between the two groups,
(3) each group independently exhibits the O--Na and C--N anticorrelations,
and (4) there are clearly distinct \ncap\ abundance patterns
in the two groups.
It is difficult to envision an unambiguous evolutionary picture 
for M22 that accounts for the entire body of observations.
Here, we illuminate this issue by comparing M22 with 
other GCs that show similar complexity, like \mbox{NGC~1851}, and
simpler GCs, like M4 and M5.\footnote{
M22 is among the more massive Milky Way GCs
 (4.0~$\times$~10$^5$~\msun,
assuming $M/L_{V} =$~2~$M_{\odot}/L_{\odot}$), and
the present-day mass of M22 is also similar to that of M4, M5, and 
\mbox{NGC~1851}
(1.2~$\times$~10$^5$~\msun,
 5.4~$\times$~10$^5$~\msun, and
 3.4~$\times$~10$^5$~\msun, respectively).}

The GCs M4 and M5 are a frequently studied pair of
clusters that are not physically related to one another.
Both are more metal-rich than M22 ([Fe/H]~$= -$1.2 and
$-$1.3), and previous work
has revealed that M4 contains moderate \spro\ enrichment
relative to M5
\citep{ivans99,ivans01,yong08a,yong08b,marino08}.
The heavy element abundances in M5 are similar to the scaled
S.S.\ \rpro\ residuals \citep{yong08a,yong08b,lai11}, and the low Pb
abundance \citep{yong08a} suggests that these elements
were produced by \rpro\ nucleosynthesis without the need to 
invoke contributions from the \spro\ \citep{roederer10b,roederer11a}. 
We subtract the heavy element abundances
in M5 from those in M4 (cf.\ \citealt{yong08b})
to estimate the \spro\ contribution to M4.
As shown in Figure~\ref{m4m5plot}, these differences are 
remarkably similar to the differences observed between the
$r+s$ and $r$-only groups in M22.
There is a gradual increase in the \spro\ content of Co through Zn
(27~$\leq Z \leq$~30),
a moderate \spro\ contribution with some element-to-element scatter
for Rb--Rh (37~$\leq Z \leq$~45),
a gradual decrease from Ba to Gd
(56~$\leq Z \leq$~64), and a gradual increase from Yb to Pb
(70~$\leq Z \leq$~82).\footnote{
Neither \citet{ivans01} nor \citet{yong08b} found differences in
[Ca/Fe] between M4 and M5.
Those studies did not examine K, and the rest of the 
abundance ratios from Ca--Mn in M4 and M5 were found to be identical.
}
There is no a priori reason to expect such similarity.
Figure~\ref{m4m5plot} implies that 
the heavy elements in M5 and the M22 $r$-only group 
were produced by similar nucleosynthesis mechanisms,
and the heavy elements in M4 and the M22 $r+s$ group 
were produced by another similar set of nucleosynthesis mechanisms.

The heavy elements in \mbox{NGC~1851} 
resemble the pattern observed in M22
\citep{yong08c,yong09,carretta10,carretta11}, and 
\citeauthor{carretta10}\ raised the possibility that \mbox{NGC~1851} 
may have formed through the merger of two proto-clusters
in a now dissolved dwarf galaxy.
The Eu abundance within each of M22 and \mbox{NGC~1851} is constant,
but moderate enhancements are observed in Zr, Ba, La, and Ce
in some stars of both GCs.
\citeauthor{carretta10}\ report a small but detectable spread in 
Fe and Ca in \mbox{NGC~1851}.
Unlike M22, these two elements are strongly correlated, which
implies the more metal-rich group is not 
enhanced in [Ca/Fe] relative to the metal-poor group.
Note, however, that \citet{lee09} suggest much larger [Ca/H] 
variations are present in \mbox{NGC~1851}.
Like M22, \mbox{NGC~1851} has a split SGB \citep{milone08}, 
which may be explained by either an age difference of
$\sim$~1~Gyr or a difference in the overall CNO with a negligible age
difference \citep{cassisi08,ventura09}.
Examination of the radial distributions of different SGB populations
gives conflicting results for \mbox{NGC~1851}, and
radial distributions for the two groups of 
stars in M22 have not been investigated.

\begin{deluxetable*}{ccccccccccc}
\tablecaption{$r$- and $s$-process Percentages in the M22 $r+s$ Group
\label{rstab}
}
\tablewidth{0pt}
\tablehead{
\colhead{Element} &
\colhead{$Z$} &
\colhead{$N_{r+s}$} &
\colhead{$N_{r}$\tablenotemark{a}} &
\colhead{$N_{s}$} &
\colhead{$\log \epsilon_{r+s}$\tablenotemark{b}} &
\colhead{$\log \epsilon_{r}$\tablenotemark{a}\tablenotemark{b}} &
\colhead{$\log \epsilon_{s}$\tablenotemark{b}} &
\colhead{\%$r$\tablenotemark{a}} &
\colhead{\%$s$} &
\colhead{$\sigma_{\%s}$}}
\startdata
Co               & 27 & 43.7     & 33.1     & 10.5                         & 3.18    & 3.06    &    2.56 &\nodata& 24.1  & $^{+11.2}_{-9.8}$  \\
Ni               & 28 & 562.     & 501.     & 61.2                         & 4.29    & 4.24    &    3.33 &\nodata& 10.9  & $^{+7.7}_{-7.1}$   \\
Cu               & 29 & 2.40     & 1.62     & 0.777                        & 1.92    & 1.75    &    1.43 &\nodata& 32.4  & $^{+11.1}_{-9.5}$  \\
Zn               & 30 & 30.2     & 19.1     & 11.1                         & 3.02    & 2.82    &    2.59 &\nodata& 36.9  & $^{+11.2}_{-9.5}$  \\
Sr               & 38 & 0.347    & 0.105    & 0.242                        & 1.08    & 0.56    &    0.92 & 30.2  & 69.8  & $^{+10.4}_{-7.7}$  \\
Y                & 39 & 0.151    & 0.0447   & 0.107                        & 0.72    & 0.19    &    0.57 & 29.5  & 70.5  & $^{+1.9}_{-1.8}$   \\
Zr~(\textsc{i})  & 40 & 0.347    & 0.151    & 0.247                        & 1.14    & 0.72    &    0.93 & 38.0  & 62.0  & $^{+6.8}_{-5.7}$   \\
Zr~(\textsc{ii}) & 40 & 0.646    & 0.302    & 0.344                        & 1.35    & 1.02    &    1.08 & 46.8  & 53.2  & $^{+7.4}_{-6.4}$   \\
Mo               & 42 & 0.0479   & 0.0468   & 0.00109                      & 0.22    & 0.21    & $-$1.42 & 97.7  & 2.3   & $^{+53.3}_{-2.3}$  \\
Ru               & 44 & 0.0513   & 0.0363   & 0.0150                       & 0.25    & 0.10    & $-$0.28 & 70.8  & 29.2  & $^{+17.7}_{-14.2}$ \\
Rh               & 45 & 0.00759  & 0.00275  & 0.00483                      & $-$0.58 & $-$1.02 & $-$0.78 & 36.3  & 63.7  & $^{+15.0}_{-10.6}$ \\
Ba               & 56 & 0.398    & 0.110    & 0.288                        & 1.14    & 0.58    &    1.00 & 27.5  & 72.5  & $^{+5.6}_{-4.6}$   \\
La               & 57 & 0.0195   & 0.00724  & 0.0122                       & $-$0.17 & $-$0.60 & $-$0.37 & 37.2  & 62.8  & $^{+2.6}_{-2.4}$   \\
Ce               & 58 & 0.0562   & 0.0155   & 0.0408                       & 0.29    & $-$0.27 &    0.15 & 27.5  & 72.5  & $^{+1.7}_{-1.6}$   \\
Pr               & 59 & 0.00661  & 0.00331  & 0.00330                      & $-$0.64 & $-$0.94 & $-$0.94 & 50.1  & 49.9  & $^{+4.0}_{-3.7}$   \\
Nd               & 60 & 0.0417   & 0.0178   & 0.0239                       & 0.16    & $-$0.21 & $-$0.08 & 42.7  & 57.3  & $^{+2.0}_{-1.9}$   \\
Sm               & 62 & 0.0126   & 0.00813  & 0.00446                      & $-$0.36 & $-$0.55 & $-$0.81 & 64.6  & 35.4  & $^{+4.8}_{-4.4}$   \\
Eu               & 63 & 0.00331  & 0.00355  & $-$0.000237\tablenotemark{c} & $-$0.94 & $-$0.91 & \nodata & 100.  & 0.0   & $^{+28.4}_{-0.0}$  \\
Gd               & 64 & 0.0182   & 0.0120   & 0.00617                      & $-$0.20 & $-$0.38 & $-$0.67 & 66.1  & 33.9  & $^{+15.0}_{-12.2}$ \\
Tb               & 65 & 0.00145  & 0.000851 & 0.000594                     & $-$1.30 & $-$1.53 & $-$1.69 & 58.9  & 41.1  & $^{+22.8}_{-16.4}$ \\
Dy               & 66 & 0.0195   & 0.0158   & 0.00365                      & $-$0.17 & $-$0.26 & $-$0.90 & 81.3  & 18.7  & $^{+13.1}_{-11.3}$ \\
Ho               & 67 & 0.00191  & 0.00191  & 0.00                         & $-$1.18 & $-$1.18 & \nodata & 100.  & 0.0   & $^{+69.8}_{-0.0}$  \\
Er               & 68 & 0.0132   & 0.00955  & 0.00363                      & $-$0.34 & $-$0.48 & $-$0.90 & 72.4  & 27.6  & $^{+17.9}_{-14.4}$ \\
Tm               & 69 & 0.000933 & 0.000741 & 0.000192                     & $-$1.49 & $-$1.59 & $-$2.18 & 79.4  & 20.6  & $^{+60.2}_{-20.6}$ \\
Yb               & 70 & 0.00708  & 0.00309  & 0.00399                      & $-$0.61 & $-$0.97 & $-$0.86 & 43.7  & 56.3  & $^{+22.3}_{-14.7}$ \\
Hf               & 72 & 0.00776  & 0.00263  & 0.00513                      & $-$0.57 & $-$1.04 & $-$0.75 & 33.9  & 66.1  & $^{+11.6}_{-8.6}$  \\
Ir               & 77 & \nodata  & 0.0324   & \nodata                      & \nodata & 0.05    & \nodata &\nodata&\nodata& \nodata	    \\    
Pb               & 82 & 0.191    & 0.0288   & 0.162                        & 0.82    & 0.00    &    0.75 & 15.1  & 84.9  & $^{+5.2}_{-3.9}$   \\
Th               & 90 & \nodata  & 0.000891 & \nodata                      & \nodata & $-$1.51 & \nodata &\nodata&\nodata& \nodata            \\
\enddata
\tablenotetext{a}{
The $r$ component implicitly includes contributions from all other processes
that may have enriched the stars in M22 prior to the epoch of \spro\ 
enrichment, e.g., charged-particle reactions, etc.}
\tablenotetext{b}{$\log\epsilon = \log N +$~1.54}
\tablenotetext{c}{Indicates mild destruction of Eu by the \spro\ (not statistically significant)}
\end{deluxetable*}

The M22 chemistry does not exclude the possibility that it
formed through a merger of two separate groups similar to M4 and M5
(at lower metallicity).
The \spro\ abundances in the two M22 groups are sufficiently distinct
that these two groups would be regarded as completely separate populations 
if not observed together in the same GC.
Similar metallicities and \rpro\ abundances might be expected
if the groups formed in close proximity
in a now-dissolved dwarf galaxy.

On the other hand, M22 shares several characteristics
with the metal-poor populations in $\omega$~Cen, 
which is more difficult to interpret as having formed
via merging of several clusters.
Based on current self-enrichment models, a possible way to account for
the M22 chemistry is through fine-tuning of
the times of accumulation of the material
from which successive generations form.
In this scenario M22 does not evolve as an isolated system,
and external gas flows can contribute 
to the enrichment processes \citep{marino11}.
A similar mechanism has been recently suggested by \citet{dantona11}
to explain the O--Na anticorrelation pattern in the more
complex case of $\omega$~Cen \citep{johnson10,marino11a}.
Further exploration is beyond the scope of the present work.

\section{The Source of the $s$-process Material}
\label{agbmass}

To summarize the results of the previous sections,
the heavy elements in the
M22 $r$-only group can be explained by nucleosynthesis mechanisms
associated with core collapse SNe.
The $r+s$ group contains a moderate amount of material produced
by \spro\ nucleosynthesis.
Previous studies have shown that the $r+s$ group has a higher mean
metallicity than the $r$-only group,
but the ratio of \rpro\ material
to Fe-group material is roughly equal in the two groups.
In this section we investigate possible nucleosynthetic sources
for the \spro\ material in the $r+s$ group.

We subtract the \rpro\ contribution (i.e., the abundance in the $r$-only
group) to each of La and Pb in the $r+s$ group
to derive the intrinsic [Pb/La]$_{s}$ ratio.
We perform a similar calculation to estimate the intrinsic
\spro\ ratios in M4 by
subtracting the M5 abundances using the \citet{yong08a} abundances.
This yields [Pb/La]$_{s} = +$0.18~$\pm$~0.09 in M22 and
[Pb/La]$_{s} = -$0.01~$\pm$~0.08 in M4.
Similarly, we derive the indices\footnote{
As defined by, e.g., \citet{bisterzo10}, the ratios of light ($ls$)
and heavy ($hs$) \spro\ yields are
[$ls$/Fe]~$= \frac{1}{2}$([Y/Fe]~$+$~[Zr/Fe]) and
[$hs$/Fe]~$= \frac{1}{3}$([La/Fe]~$+$~[Nd/Fe]~$+$~[Sm/Fe]).
Also, [$hs$/$ls$]~$=$~[$hs$/Fe]~$-$~[$ls$/Fe].  
}
[$hs$/$ls$]$_{s}= -$0.01 and $-$0.50 and
[Pb/$hs$]~$= +$0.29 and $+$0.28 for M22 and M4, respectively.
(Uncertainties on each of these quantities are likely 
0.10--0.15~dex.)
These ratios and indices are useful since they are insensitive to the 
dredge-up efficiency or the dilution of AGB products in the
stars currently observed.
We infer that the AGBs providing the \spro\ enrichment
in M4 and the $r+s$ group in M22 
were similar but not identical.

Models of \spro\ nucleosynthesis 
indicate that Pb is a sensitive probe of 
the stellar mass, metallicity, and neutron flux.
\citet{goriely00} present yields for a model representative of
1.5~$\leq M \leq$~3.0~\msun, [Fe/H]~$= -$1.25 AGB stars,
and [Pb/La] can be estimated from Figure~3 of \citet{goriely01} 
for their 3~\msun\ zero-metallicity AGB model.
\citet{cristallo09} present yields for 2~\msun\ AGB models
at [Fe/H]~$= -$1.2 and $-$2.2.
\citet{bisterzo10} present a set of yields 
for several masses
($M =$~1.3, 1.4, 1.5, and 2.0~\msun), 
metallicities ($-$3.6~$\leq$~[Fe/H]~$\leq -$1.0), 
and $^{13}$C pocket efficiencies.
[Pb/La] predictions can also be calculated for limited combinations of
masses, metallicities, and $^{13}$C pocket efficiencies from the
AGB yields presented in \citet{roederer10b}.
These predictions rely on
similar atomic data, stellar models, assumptions
about branching points, etc., 
and so are not entirely independent.

When compared with these yields,
the M4 and M22 \spro\
heavy element ratios and indices point to a common theme: 
low mass AGB stars ($M \leq$~3~\msun) cannot reproduce the observed values
unless the standard $^{13}$C pocket efficiency 
is reduced by factors of 30--150.
Pb is enhanced in both M4 and the $r+s$ group in M22
relative to the lighter \ncap\ elements and Fe, but
it is not nearly as enhanced as observed in metal-poor stars
extrinsically enriched in \spro\ elements by an AGB binary companion.
AGBs with $M \sim$~4.5--6.0~\msun\ (those which may not
form a $^{13}$C pocket and hence will not activate the 
$^{13}$C($\alpha$,$n$)$^{16}$O neutron source)
can produce lower [Pb/La] ratios \citep{roederer10b}.
For comparison, predictions for the 
5~\msun\ AGB models at [Fe/H]~$= -$2.3
are shown in Figure~\ref{gcpbplot}.
Figures in \citet{bisterzo10} present the [$hs$/$ls$]
and [Pb/$hs$] indices for a limited number of 3 and 5~\msun\ AGB models.
Their predictions for the appropriate (low) $^{13}$C pocket efficiency
in a 5~\msun\ AGB
are a near-perfect match to the \spro\ ratios in each of M4 and M22
at their respective metallicities.
This result is encouraging.

The $^{22}$Ne neutron source, which activates at higher temperatures than 
the $^{13}$C neutron source, does not play
a dominant role in AGB stars with $M <$~3--4~\msun.
In AGB stars with $M =$~5--8~\msun, the temperature at the base of the thermal 
pulse is
higher, and the $^{22}$Ne($\alpha$,$n$)$^{25}$Mg reaction can occur there
(e.g., \citealt{busso01}).
In principle
this could also account for the \spro\ neutron captures that produce
small amounts of Co--Zn, as observed; see
\citet{yong08b} and \citet{karakas09} for further discussion.

Models of the weak component of the \spro\ have traditionally
been set in $\sim$~25~\msun\ stars (e.g., \citealt{raiteri93})
that activate the $^{22}$Ne neutron source during core He-burning
and shell C-burning stages, since models of less massive stars suggest
that subsequent burning stages will destroy any \spro\ material created.
Models that include rotationally-induced mixing can increase
the neutron flux by mixing $^{14}$N (which is converted to $^{22}$Ne)
into the relevant regions, possibly producing nuclei as
heavy as $^{208}$Pb \citep{pignatari08}.
Yet the enhanced \spro\ abundances observed in the M22 $r+s$ group
cannot be due to the operation of the weak \spro\ in massive stars.
There is no reason to expect that the the SNe
that enriched the metal-rich $r+s$ group 
in M22 host the weak \spro\ and those that enriched the
metal-poor $r$-only group did not.

The minority neutron-rich Mg isotopes $^{25}$Mg and $^{26}$Mg
may be produced (among other proton- and $\alpha$-capture channels) 
by the reaction sequence
$^{22}$Ne($\alpha$,$n$)$^{25}$Mg($n$,$\gamma$)$^{26}$Mg, which
acts as both a neutron source and poison.
Preliminary measurements of
($^{25}$Mg$+^{26}$Mg)/$^{24}$Mg 
in M4 and M5 indicate that the Mg isotopes have similar proportions
in the two clusters \citep{yong08b}.
While preliminary, these measurements hint that the source affecting
the Mg isotopic ratios has acted similarly in M4 and M5.  
Since moderate quantities of \spro\ material 
are observed in M4 and M22 but not M5, it 
seems unlikely that the source of the \spro\ material modifies the Mg
isotopic ratios substantially.  
Unfortunately we cannot assess the Mg isotopic ratios from our M22 data, 
but new measurements of these ratios 
in all three GCs would be of great interest.

At low metallicity
$^{22}$Ne also serves as a primary seed nucleus
from which a chain of \ncap\ reactions can generate a 
small leakage across the Fe-group isotopes \citep{busso01,gallino06}.
We propose that the observed variations in the Fe-group ratios
and perhaps even the overall
metallicity (Fe) increase in the $r+s$ group 
could be due to this phenomenon.
The \spro\ path passes through stable or
long-lived nuclei of K, Ca, Ti, V, and Cr,
including several nuclei 
($^{39}$K, $^{42}$Ca, $^{43}$Ca, $^{44}$Ca, $^{50}$Ti, 
$^{51}$V, $^{52}$Cr) with closed nuclear shells.
The only stable isotopes of Sc and Mn, $^{45}$Sc and $^{55}$Mn, 
do not have closed nuclear shells, 
so it is perhaps surprising that Sc shows an enhancement 
while Mn shows a deficiency in the $r+s$ group.
The fact that we observe no change in Ti or Cr 
could be related to the initially larger abundances
of these even-$Z$ elements relative to a small \spro\ contribution.
Ca, which could also be expected to follow this pattern,
may be enhanced because there are three Ca isotopes on 
the \spro\ path with closed proton shells.
Obviously detailed calculations 
are needed to test these proposals for the
Fe-group variations between the two groups in M22.

\begin{figure}
\begin{center}
\includegraphics[angle=0,width=3.3in]{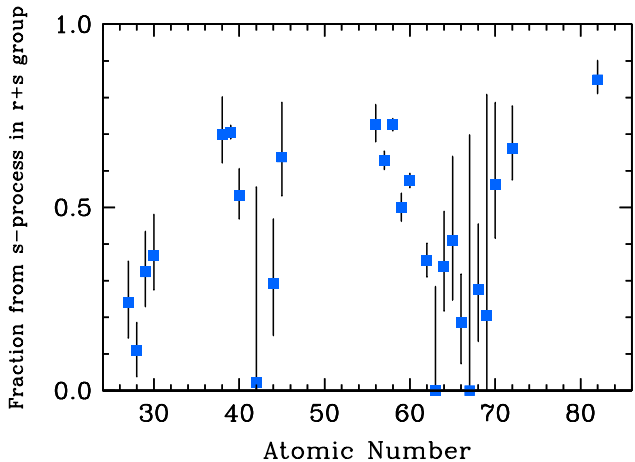}
\end{center}
\caption{
\label{spercentplot}
The fractional component of the $r+s$ group of stars originating
in the \spro\ for elements heavier than Fe.
}
\end{figure}

If the \spro\ material in M4 and M22 is produced by 
neutrons from the $^{22}$Ne source,
this implies an origin different from that of the \spro\ material in
$\omega$~Cen.
\citet{smith00} found that the \ncap\ elements in $\omega$~Cen are best fit
by low-mass (1.5--3.0~\msun) AGB stars where the $^{13}$C neutron
source is active.
The observed [Rb/Zr] ratios in $\omega$~Cen, 
which are quite sensitive to the
neutron density and hence the neutron source
because of \spro\ branching at $^{85}$Kr,
are best fit by low mass ($M \leq$~3~\msun) AGB models.
The [Rb/Zr] ratios in M4 derived by \citet{yong08a} 
are higher than those in $\omega$~Cen, 
and our [Rb/Zr] ratio in the M22 $r+s$ group
is not lower than that in M4,
supporting our assertion. 
(Recall that we could only derive upper limits on the Rb abundance in M22.)
Furthermore, \citet{cunha02} found no evolution in the [Cu/Fe] ratio 
over $-$2.0~$<$~[Fe/H]~$< -$0.8 in $\omega$~Cen,
indicating that there were no contributions to Cu from 
AGB stars that could produce Cu with neutrons from
the $^{22}$Ne($\alpha$,$n$)$^{25}$Mg reaction.
\citet{dantona11} point out that the timescales for establishing
the light element variations and the \spro\ enrichement 
in $\omega$~Cen are discrepant, and this issue is not yet resolved.
 
These constraints raise an obvious question: if the \spro\ material
in M4 and M22
is produced in more massive AGB stars, then
why is \spro\ material not detected
in \textit{every} cluster where the light element
variations are observed?
\citet{marino08} showed that there might be a weak correlation
between [Ba/Fe] and [Al/Fe] in M4, 
a point also investigated by \citet{smith08}.
Those data also suggest weak correlations between [Ba/Fe] and each of
[Na/Fe] and [Si/Fe].
For the majority of GCs, however, such correlations are not found
(e.g., \citealt{armosky94,dorazi10a}).
This supports our conclusion, drawn from the [Pb/Eu] ratios, that
the heavy elements in most metal-poor GCs are produced
by \rpro\ nucleosynthesis.
Perhaps in GCs like M4, the $r+s$ group in M22, or the metal-rich group
of \mbox{NGC~1851},
material from slightly lower AGB masses
was allowed to enrich the GC ISM before the clusters formed.
This might suggest that these particular clusters 
were more massive initially or originated in dwarf galaxies
whose potentials could more easily retain ejecta
and sustain extended periods of star formation.
This scenario is appealing because several of the
metal-poor clusters exhibiting \spro\ enrichment (M22, \mbox{NGC~1851}, 
$\omega$~Cen) exhibit at least minimal spreads in Fe
and (in the case of M22 and \mbox{NGC~1851})
could have been formed through mergers.

\begin{figure}
\begin{center}
\includegraphics[angle=0,width=3.3in]{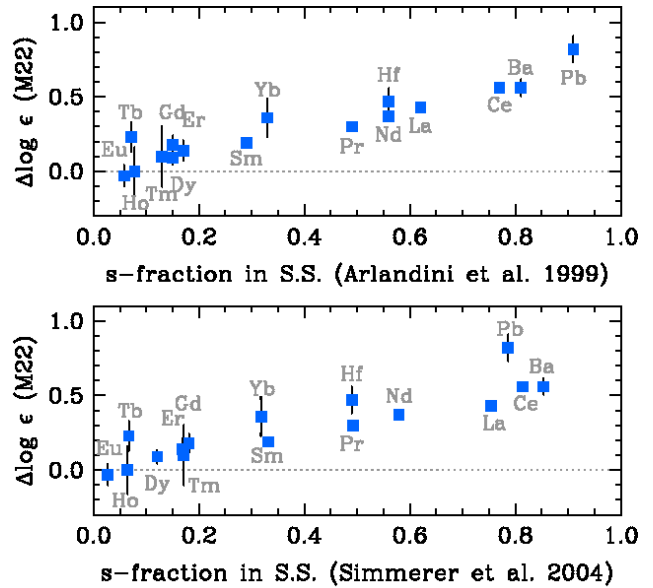}
\end{center}
\caption{
\label{sfracplot}
Differences in the mean abundances between the $r$-only group
and the $r+s$ group as a function of the $s$-fraction of 
each element in the S.S.
The top panel shows the $s$-fraction as calculated from the
average yields of the 1.5 and 3.0~\msun\ stellar models at [Fe/H]~$= -$0.3
of \citet{arlandini99}, including the contribution of
low-metallicity AGB stars to the S.S.\ Pb as
derived by \citet{travaglio01}.
The bottom panel shows the $s$-fraction as calculated by the
classical method \citep{simmerer04}.
The dotted lines indicate zero difference.
}
\end{figure}

In summary, the observed \spro\ abundance patterns are not well-fit
by low-metallicity models of AGB stars with $M \leq$~3~\msun.
Higher mass AGBs that activate the $^{22}$Ne neutron source
may provide a better fit. 
Both stellar groups in M22 exhibit the Na--O anticorrelation, 
but the observed lack of a correlation between \spro\ 
enrichment and Na within the $r+s$ group 
is difficult to understand if these elements are all produced 
by AGB stars of higher masses.
We encourage more detailed 
exploration of the possible association between the \spro\ products
in these metal-poor GCs with intermediate-mass AGB stars.

\section{An Empirical $s$-process Abundance Distribution}
\label{spro}

In this section we compare the nature of low-metallicity
\spro\ enrichment with the \spro\ abundance pattern observed in the S.S.
The M22 \spro\ ``residual'' is derived by subtracting the 
abundances in the $r$-only group from the abundances in the $r+s$ group.
This method assumes that the \rpro\ material in both groups is identical,
as indicated by observations.

Table~\ref{rstab} lists the abundances and $r$- and \spro\ 
fractions for the heavy elements in M22.
Figure~\ref{spercentplot} illustrates the fraction of each of these
elements that originates in the \spro\ in the M22 $r+s$ group.
Elements on the \spro\ path with closed neutron shells 
(Sr--Zr, Ba--Nd, and Pb), 
plus a few others (Rh, Yb, Hf), 
owe more than 50\% of their
abundance in the $r+s$ group to the \spro.
Elements in the middle of the rare earth domain 
(Sm--Tm) and just beyond the first \spro\ peak (Mo, Ru) 
are still mostly made of \rpro\ material, with \spro\ fractions
less than 40\% or so.
More than 80\% of the Pb in the $r+s$ group originated in the \spro,
the most of any element studied.
Several elements, including Mo, Eu, Ho, and Tm, are consistent
with a pure \rpro\ origin 
(i.e., show no enhancement in the $r+s$ group)
within the uncertainties.
In analogy with S.S.\ $r$-residuals derived via the classical approach,
elements with small \spro\ fractions have the 
largest \spro\ fraction uncertainties,
and elements with large \spro\ fractions have the smallest uncertainties.

To compare the \spro\ fractions in M22 with the \spro\ fractions
derived from the stellar model and classical approach, 
Figure~\ref{sfracplot} displays the elemental abundance differences
between the two M22 groups as a function of the \spro\ fraction
in the S.S.
Only elements produced predominantly by the main and strong \spro\ components
in the S.S.\ are shown (i.e., $Z \geq$~56),
since the yields of these elements should be less
sensitive to the source of the neutron flux.
There is a remarkably clear correlation, which changes
little when different stellar model \spro\ fractions (e.g.,
\citealt{arlandini99,bisterzo10}) or classical method \spro\ 
fractions (e.g., \citealt{burris00,simmerer04}) are used.
Note that the \spro\ fraction of Pb shown in Figure~\ref{sfracplot}
accounts for the low-metallicity AGB component according
to the Galactic chemical evolution model of \citet{travaglio01}.
The stellar model of \citet{gallino98} and \citet{arlandini99} 
designated the standard case for the mass of the $^{13}$C pocket
as that which best reproduced the 
S.S.\ main \spro\ component in low-mass (1.5 and 3.0~\msun)
AGB models with [Fe/H]~$= -$0.3.
The \spro\ material in the S.S.\ was produced
by a variety of AGB sources over many Gyr.
Despite this fact,
Figure~\ref{sfracplot} suggests that---at least for the 
elements with 56~$\leq Z \leq$~72---the relative yields
of low-metallicity, higher-mass AGB stars are not that different
from the more metal-rich, lower-mass AGB stars.

In the M22 $r+s$ group, 
62\% of the total amount of $Z \geq$~38
elements examined (excluding Ir and Th)
originated in the \spro.
The \spro\ contributes 79\% of the material to these same elements in the S.S.\
\citep{sneden08}.
Hypothetically, if one wants to further enrich the heavy elements in the 
M22 $r+s$ group to match the S.S.\ abundances,
a greater fraction of \spro\ material (with respect to \rpro\ material) 
needs to be added. 
In principle, then, these data support the general understanding that
\spro\ enrichment occurs at later times than \rpro\ enrichment.

\section{Conclusions}
\label{conclusions}

One longstanding obstacle to properly interpret
models of the \spro\ is having observations of pure \spro\
material outside the S.S.\ to compare with, especially since
nearly all stars contain at least a trace of \rpro\ material.
Here we provide one solution to this problem by deriving the
abundance patterns in two related groups of stars
in the metal-poor GC M22.
One group shows an \rpro\ pattern 
with no detectable enrichment by \spro\ material (the $r$-only group),
while the other group shows an additional \spro\ enhancement
(the $r+s$ group).
By subtracting the \rpro\ abundance pattern of the former from the
$r+s$ abundance pattern in the latter, we explicitly remove the \rpro\ 
contribution to reveal the \spro\ ``residual.''

The \spro\ abundance pattern in M22
strongly disfavors low mass ($M \leq$~3~\msun), low-metallicity 
AGB models.
Although no published model results span the 
appropriate range of AGB masses at the metallicity of M22,
the limited predictions available for more massive AGB stars
at low metallicity fit the data better,
especially the moderate Pb enhancement.
Predictions for $M =$~4.5 and 5.0~\msun\ AGB models at [Fe/H]~$= -$1.6
and $-$2.3 do fit the M22 \spro\ abundances, 
although 3~$< M <$~4.5~\msun\
models cannot be excluded because no predictions are available.
The neutrons that fuel the \spro\ in these models mainly
originate in the $^{22}$Ne($\alpha$,$n$)$^{25}$Mg reaction,
which requires higher activation temperatures than the 
$^{13}$C($\alpha$,$n$)$^{16}$O reaction.
In principle this could explain observed overabundances of K, Ca, Sc, 
V, Co, Ni, Cu, and Zn in the $r+s$ group.
We also calculate the $r$- and \spro\ fractions of each \ncap\ element.
This approach assumes nothing about the $r$- and \spro\ fractions
in S.S.\ material.
We encourage investigations of \spro\ nucleosynthesis 
in models with the appropriate metallicity and AGB mass range to 
better understand the origin of the heavy elements in M22.
More generally, we hope that these data will serve as useful benchmarks for
modeling and interpreting \spro\ abundance patterns and enrichment
in low metallicity environments.

Furthermore, these abundances can help interpret the
enrichment history of M22.
The $Z \geq$~27 abundance pattern in the M22 $r$-only and $r+s$ groups 
bear striking resemblance to the
(physically unrelated) GCs M5 and M4, respectively.
The $r+s$ group in M22 may share a similar 
enrichment history to M4 and possibly 
the metal-rich group in \mbox{NGC~1851}.
If the \spro\ in M22 did originate in more massive AGB stars,
this places strong constraints on the timescale for chemical enrichment,
particularly in attempting to explain why the majority of metal-poor
GCs \textit{do not} show similar signatures of \spro\ enrichment.
AGB models that can simultaneously explain the observed abundance
patterns resulting from both proton- and neutron-capture reactions
(the light element variations and \spro\ enrichment)
should prove enlightening in this regard.

\acknowledgments

We thank the referee, S.\ Cristallo, for providing a helpful
report on this work, and we also appreciate comments from
G.\ Preston on an earlier version of the manuscript.
I.U.R.\ is supported by the Carnegie Institution of Washington 
through the Carnegie Observatories Fellowship.
C.S.\ is supported by the U.S.\ National Science Foundation 
(grant AST~09-08978).

{\it Facilities:} 
\facility{Magellan:Clay (MIKE)}

\appendix
\section{Line-by-Line Mean Offsets}
\label{appendix}

\begin{deluxetable*}{ccccccccccc}
\tablecaption{Line-by-line Mean Offsets
\label{appendixtab}
}
\tablewidth{0pt}
\tablehead{
\colhead{Species} &
\colhead{$\lambda$ (\AA)} &
\colhead{$\langle\Delta\rangle$} &
\colhead{$\sigma$} &
\colhead{$\sigma_{\mu}$} &
\colhead{Species} &
\colhead{$\lambda$ (\AA)} &
\colhead{$\langle\Delta\rangle$} &
\colhead{$\sigma$} &
\colhead{$\sigma_{\mu}$} }
\startdata
Y~\textsc{ii}  & 4883.68 &  $-$0.009                      & 0.081 & 0.033  &   Nd~\textsc{ii} & 4021.33 &  $-$0.081 & 0.108 & 0.022 \\                        
Y~\textsc{ii}  & 4982.13 &  $-$0.003                      & 0.086 & 0.035  &   Nd~\textsc{ii} & 4059.95 &  $-$0.095 & 0.144 & 0.029 \\                        
Y~\textsc{ii}  & 5087.42 &  0.005                         & 0.076 & 0.031  &   Nd~\textsc{ii} & 4232.37 &  $-$0.034 & 0.074 & 0.015 \\                        
Y~\textsc{ii}  & 5119.11 &  0.007                         & 0.046 & 0.019  &   Nd~\textsc{ii} & 4446.38 &  $-$0.013 & 0.051 & 0.010 \\                        
Y~\textsc{ii}  & 5200.41 &  $-$0.003                      & 0.051 & 0.021  &   Nd~\textsc{ii} & 4462.98 &  0.216    & 0.227 & 0.046 \\                        
Y~\textsc{ii}  & 5205.73 &  0.034                         & 0.045 & 0.018  &   Nd~\textsc{ii} & 4465.06 &  0.054    & 0.067 & 0.014 \\                        
Y~\textsc{ii}  & 5289.82 &  $-$0.032                      & 0.095 & 0.039  &   Nd~\textsc{ii} & 4465.59 &  $-$0.006 & 0.051 & 0.010 \\                        
Zr~\textsc{ii} & 4050.33 &  $-$0.200                      & 0.230 & 0.163  &   Nd~\textsc{ii} & 4501.81 &  0.020    & 0.069 & 0.014 \\                        
Zr~\textsc{ii} & 4613.92 &  0.145                         & 0.180 & 0.127  &   Nd~\textsc{ii} & 4567.61 &  0.002    & 0.048 & 0.010 \\                        
Zr~\textsc{ii} & 5112.28 &  0.055                         & 0.066 & 0.047  &   Nd~\textsc{ii} & 4645.76 &  $-$0.012 & 0.048 & 0.010 \\                        
La~\textsc{ii} & 3988.51 &  $-$0.216                      & 0.230 & 0.064  &   Nd~\textsc{ii} & 4706.54 &  0.068    & 0.075 & 0.015 \\                        
La~\textsc{ii} & 3995.74 &  $-$0.115                      & 0.185 & 0.051  &   Nd~\textsc{ii} & 4797.15 &  $-$0.098 & 0.141 & 0.029 \\                        
La~\textsc{ii} & 4086.71 &  $-$0.076                      & 0.130 & 0.036  &   Nd~\textsc{ii} & 4825.48 &  0.007    & 0.041 & 0.008 \\                        
La~\textsc{ii} & 4322.50 &  $-$0.085                      & 0.098 & 0.027  &   Nd~\textsc{ii} & 4859.03 &  0.023    & 0.042 & 0.009 \\                        
La~\textsc{ii} & 4662.50 &  0.000                         & 0.031 & 0.009  &   Nd~\textsc{ii} & 4902.04 &  0.084    & 0.091 & 0.019 \\                        
La~\textsc{ii} & 4748.73 &  $-$0.083                      & 0.093 & 0.026  &   Nd~\textsc{ii} & 4914.38 &  $-$0.008 & 0.023 & 0.005 \\                        
La~\textsc{ii} & 4804.04 &  0.068                         & 0.074 & 0.021  &   Nd~\textsc{ii} & 5089.83 &  $\equiv$0.0\tablenotemark{a}  & 0.065 & 0.013 \\   
La~\textsc{ii} & 4920.98 &  0.161                         & 0.172 & 0.048  &   Nd~\textsc{ii} & 5092.79 &  $-$0.024 & 0.059 & 0.012 \\                        
La~\textsc{ii} & 4986.82 &  0.034                         & 0.071 & 0.020  &   Nd~\textsc{ii} & 5130.59 &  $-$0.059 & 0.073 & 0.015 \\                        
La~\textsc{ii} & 5114.56 &  0.089                         & 0.100 & 0.028  &   Nd~\textsc{ii} & 5132.33 &  $-$0.055 & 0.102 & 0.021 \\                        
La~\textsc{ii} & 5290.84 &  $-$0.074                      & 0.097 & 0.027  &   Nd~\textsc{ii} & 5234.19 &  $-$0.034 & 0.047 & 0.010 \\                        
La~\textsc{ii} & 5303.53 &  0.068                         & 0.092 & 0.026  &   Nd~\textsc{ii} & 5249.58 &  0.070    & 0.083 & 0.017 \\                        
La~\textsc{ii} & 6262.29 &  0.093                         & 0.100 & 0.028  &   Nd~\textsc{ii} & 5255.51 &  $-$0.006 & 0.038 & 0.008 \\                        
La~\textsc{ii} & 6390.48 &  0.136                         & 0.145 & 0.040  &   Nd~\textsc{ii} & 5293.16 &  $-$0.052 & 0.090 & 0.018 \\                        
La~\textsc{ii} & 6774.27 &  $\equiv$0.0\tablenotemark{a}  & 0.075 & 0.021  &   Nd~\textsc{ii} & 5319.81 &  0.035    & 0.047 & 0.010 \\                        
Ce~\textsc{ii} & 4073.47 &  $-$0.082                      & 0.108 & 0.028  &   Sm~\textsc{ii} & 4318.93 &  $-$0.034 & 0.067 & 0.024 \\                        
Ce~\textsc{ii} & 4083.22 &  0.046                         & 0.087 & 0.023  &   Sm~\textsc{ii} & 4434.32 &  0.053    & 0.083 & 0.029 \\                        
Ce~\textsc{ii} & 4120.83 &  0.043                         & 0.056 & 0.014  &   Sm~\textsc{ii} & 4467.34 &  $-$0.124 & 0.137 & 0.048 \\                        
Ce~\textsc{ii} & 4127.36 &  $-$0.180                      & 0.204 & 0.053  &   Sm~\textsc{ii} & 4536.51 &  0.038    & 0.055 & 0.019 \\                        
Ce~\textsc{ii} & 4137.65 &  $-$0.053                      & 0.146 & 0.038  &   Sm~\textsc{ii} & 4537.94 &  $-$0.086 & 0.091 & 0.032 \\                        
Ce~\textsc{ii} & 4222.60 &  0.027                         & 0.100 & 0.026  &   Sm~\textsc{ii} & 4591.81 &  0.023    & 0.086 & 0.030 \\                        
Ce~\textsc{ii} & 4364.65 &  $-$0.073                      & 0.100 & 0.026  &   Sm~\textsc{ii} & 4642.23 &  0.075    & 0.089 & 0.031 \\                        
Ce~\textsc{ii} & 4418.78 &  0.041                         & 0.056 & 0.014  &   Sm~\textsc{ii} & 4669.64 &  $-$0.041 & 0.061 & 0.022 \\                        
Ce~\textsc{ii} & 4486.91 &  $-$0.023                      & 0.112 & 0.029  &   Sm~\textsc{ii} & 4719.84 &  0.096    & 0.110 & 0.039 \\                        
Ce~\textsc{ii} & 4560.96 &  0.016                         & 0.045 & 0.012  &   Eu~\textsc{ii} & 3907.11 &  $-$0.246 & 0.252 & 0.178 \\                        
Ce~\textsc{ii} & 4562.36 &  0.016                         & 0.046 & 0.012  &   Eu~\textsc{ii} & 4129.72 &  0.009    & 0.087 & 0.062 \\                        
Ce~\textsc{ii} & 4572.28 &  0.037                         & 0.075 & 0.019  &   Eu~\textsc{ii} & 6645.06 &  0.237    & 0.258 & 0.182 \\                        
Ce~\textsc{ii} & 4582.50 &  0.055                         & 0.095 & 0.025  &   Gd~\textsc{ii} & 4130.37 &  0.173    & 0.190 & 0.134 \\                        
Ce~\textsc{ii} & 4628.16 &  0.089                         & 0.102 & 0.026  &   Gd~\textsc{ii} & 4251.73 &  $-$0.097 & 0.152 & 0.108 \\                        
Ce~\textsc{ii} & 5274.23 &  0.032                         & 0.063 & 0.016  &   Gd~\textsc{ii} & 4498.29 &  $-$0.076 & 0.127 & 0.090 \\                        
Ce~\textsc{ii} & 5330.56 &  0.011                         & 0.087 & 0.022  &   Dy~\textsc{ii} & 3694.81 &  $-$0.108 & 0.321 & 0.185 \\                        
Pr~\textsc{ii} & 4222.95 &  0.012                         & 0.041 & 0.024  &   Dy~\textsc{ii} & 3983.65 &  0.073    & 0.120 & 0.069 \\                        
Pr~\textsc{ii} & 4408.81 &  $-$0.041                      & 0.050 & 0.029  &   Dy~\textsc{ii} & 4073.12 &  $-$0.078 & 0.153 & 0.088 \\                        
Pr~\textsc{ii} & 5259.73 &  0.012                         & 0.051 & 0.030  &   Dy~\textsc{ii} & 4449.70 &  0.113    & 0.161 & 0.093 \\                        
Pr~\textsc{ii} & 5322.77 &  0.017                         & 0.024 & 0.014  &                  &         &           &       &       \\
\enddata
\tablecomments{
For a given line, 
the mean offset $\langle\Delta\rangle$ is computed 
as the average over all 6~stars of the offset
relative to the mean of all other lines of the same 
species in a given star.
Exceptions are Gd~\textsc{ii}, whose mean is computed
without \mbox{I-80}, and Dy~\textsc{ii}, whose mean
is computed without \mbox{I-27} and \mbox{I-80}.
}
\tablenotetext{a}{
The $\log(gf)$ values for these two lines are not given
in the literature and are derived here to empirically 
match the mean abundance derived from other lines 
in each of the 6 stars.
See text for details.
}
\end{deluxetable*}

We have calculated line-by-line mean offsets for \ncap\ species
whose abundance is derived from three or more lines.
Such information is useful when comparing abundances 
from different studies that use a small number of 
non-overlapping lines.
In Table~\ref{appendixtab}, we list the species (columns~1 and 6), 
wavelength (columns~2 and 7), average offset from the mean abundance
as derived for each of the 6~stars examined (columns~3 and 8),
standard deviation of the average offset (columns~4 and 9), 
and standard deviation of the mean of the average offset (columns~5 and 10).

Because the number of La~\textsc{ii} and Nd~\textsc{ii} lines
examined is large, we have also derived empirical 
$\log(gf)$ values for two lines not covered in the
\citet{lawler01a} and \citet{denhartog03} laboratory studies,
La~\textsc{ii} 6774.27\AA\ ($\log(gf) = -$1.77~$\pm$~0.06)
and
Nd~\textsc{ii} 5089.83\AA\ ($\log(gf) = -$1.27~$\pm$~0.06).
These lines are not used in determining the abundances in M22.
This La~\textsc{ii} line is often one of the only lines available
in studies that target the red region of the spectrum.
\citet{johnson10} provide empirical corrections to the abundance
to account for the unknown HFS pattern of this line.
In the M22 stars observed, the EWs of this line are all
10--20~m\AA, so the correction is approximately zero.

\end{document}